\pdfoutput=1
\documentclass[reprint,superscriptaddress]{revtex4-2}

\usepackage[T1]{fontenc} 
\usepackage[utf8x]{inputenc}
\DeclareUnicodeCharacter{2212}{-}

\usepackage[ngerman]{datetime}
\newdateformat{myformat}{\THEDAY{. }\monthnamengerman[\THEMONTH] \THEYEAR}
\usepackage{parskip} 
\usepackage{svg}
\usepackage{import}
\usepackage{color}
\usepackage{float}
\usepackage{subfig}
\usepackage{placeins}
\usepackage{tikz}
\usepackage{pgfplots}
\usepgfplotslibrary{external} 
 \tikzset{external/system call ={lualatex
 \tikzexternalcheckshellescape
 -halt-on-error
 -interaction=batchmode
 -jobname "\image" "\textsource"}}
\usepackage{amsmath} 
\usepackage{amssymb} %
\usepackage{wasysym} %
\usepackage{marvosym}
\usepackage{mathtools}
\usepackage{upgreek}
\usepackage{xcolor}
\usepackage{paralist}
\usepackage{enumitem} 
\usepackage{xcolor,hyperref}
\hypersetup{ 
	colorlinks=true,       
    linkcolor=blue,          
    citecolor=blue,        
    filecolor=magenta,      
    urlcolor=cyan           
    }
\usepackage[format=plain,
      justification=RaggedRight]
     {caption}
\newcommand{\subR}[1]{
  \protect\begin{NoHyper}\protect\subref{#1}\protect\end{NoHyper}
}

\newcommand{\R}{\mathbb{R}}
\newcommand{\Z}{\mathbb{Z}}

\newcommand{\mathsym}[1]{{}}

\newcommand{\vc}[1]{\boldsymbol{#1}}

\newcommand{\bra}[1]{\left( #1 \right)}
\newcommand{\brc}[1]{\left[ #1 \right]}
\newcommand{\bro}[1]{\left\{ #1 \right\}}

\newcommand{\g}[1]{\begin{gather}      #1     \end{gather}}

\newcommand{\alg}[1]{\begin{align}      #1     \end{align}}

\begin{document}

\title{Ensnarled: On the topological linkage of spatially embedded network pairs}
\author{Felix Kramer}
\email[]{felixuwekramer@proton.me}
\affiliation{Max Planck Institute for Molecular Cell Biology and Genetics (MPI-CBG),  Dresden 01307, Germany}
\affiliation{Center for Systems Biology Dresden (CSBD),  Dresden 01307, Germany}
\author{Carl D. Modes} 
\email[]{modes@mpi-cbg.de}
\affiliation{Max Planck Institute for Molecular Cell Biology and Genetics (MPI-CBG), Dresden 01307, Germany}
\affiliation{Center for Systems Biology Dresden (CSBD),  Dresden 01307, Germany}
\affiliation{Cluster of Excellence, Physics of Life, TU  Dresden, Dresden 01307, Germany}
\date{\today}
\begin{abstract}
The observation, design and analysis of mesh-like networks in bionics, polymer physics and biological systems has brought forward an extensive catalog of fascinating structures of which a subgroup share a particular, yet critically under appreciated attribute: being embedded in space such that one wouldn't be able to pull them apart without prior removal of a subset of edges, a state which we here call \textit{ensnarled}.
In this study we elaborate on a graph theoretical method to analyze ensnarled finite, 2-component nets on the basis of Hopf-link identification.
Doing so we are able to construct an \textit{edge priority} operator $\vc{\Lambda}$, derived from the linking numbers of the spatial graphs' cycle bases, which highlights critical edges.
On its basis we developed a greedy algorithm which identifies optimal edge removals to achieve unlinking, allowing for the establishment of a new topological metric characterizing the state of ensnarled network pairs.
\end{abstract}

\maketitle
\texttt{\section{Introduction}
\label{sec:intro}}
How can one quantify the degree to which two networks topologically constrain each other?
This question emerges throughout biology, chemistry and physics, e.g.  when considering coiled DNA~\cite{RN415}, compounds of vasculature~\cite{RN168,RN400} or multi-component crystals~\cite{RN399,RN381,RN390,RN416,RN380,RN401}.
It becomes particularly tricky when one has to distinguish two such objects from each other, for example by the number of cuts one has to perform to split the components.
Yet, most work here has been performed on the characteristics of individual closed curves or intrinsic linking of one-component graph embeddings~\cite{RN405,RN367}.
While partial generalizations have been performed for extended meshes (and their graphical representations~\cite{RN403,RN409}) much remains to be done for spatially embedded networks in 3D, where the crossing of edges is costly if not physically impossible.
As this property has no proper name, we suggest \textit{ensnarled} and present here the means of quantification.
We find to our best knowledge that no attempt has been made to characterize such amorphous, tubal structures for being ensnarled with their environment, a phenomenon particularly present in multi-component biological transport systems, e.g. vasculature.
The mammalian liver lobule provides an illustrative example: here, intricate, intertwined sinusoid (vascular) and bile canalicular networks are allowed a significant spatial overlap with mediating cells in between, enabling the actual filtering function of the organ~\cite{RN3,RN152}.
This system has further been in the focus of a multitude of studies concerning matters of non-trivial morphogenesis and network robustness~\cite{luminaExpansion,RN26,RN167,RN175}.
Yet no useful analytic measure of it being ensnarled exist.\newline
We develop a graph-theoretic framework which allows the identification of critical edges involved in the linkage of two spatially embedded networks.
To achieve these means we introduce a linear operator, ${\vc{\Lambda}}$, which maps any two cycles (in edge-space representation) on to their respective linking number.
This operator ${\vc{\Lambda}}$ quantifies a new form of edge priority, i.e. the importance of edges for the topological linkage of the system.
We demonstrate that this allows for the efficient identification of minimal cuts between two linked structures (which should not be confused with the approach taken in percolation theory).
We go on to survey various network archetypes and conclude by discussing potential implications and applications of~${\vc{\Lambda}}$.
 \\

\section{Theory}
\label{sec:theory}
\begin{figure} [t]
\centering
	\subfloat[]{
	\def\svgscale{0.3}
\begingroup%
  \makeatletter%
  \providecommand\color[2][]{%
    \errmessage{(Inkscape) Color is used for the text in Inkscape, but the package 'color.sty' is not loaded}%
    \renewcommand\color[2][]{}%
  }%
  \providecommand\transparent[1]{%
    \errmessage{(Inkscape) Transparency is used (non-zero) for the text in Inkscape, but the package 'transparent.sty' is not loaded}%
    \renewcommand\transparent[1]{}%
  }%
  \providecommand\rotatebox[2]{#2}%
  \newcommand*\fsize{\dimexpr\f@size pt\relax}%
  \newcommand*\lineheight[1]{\fontsize{\fsize}{#1\fsize}\selectfont}%
  \ifx\svgwidth\undefined%
    \setlength{\unitlength}{255.50312805bp}%
    \ifx\svgscale\undefined%
      \relax%
    \else%
      \setlength{\unitlength}{\unitlength * \real{\svgscale}}%
    \fi%
  \else%
    \setlength{\unitlength}{\svgwidth}%
  \fi%
  \global\let\svgwidth\undefined%
  \global\let\svgscale\undefined%
  \makeatother%
  \begin{picture}(1,1.41198504)%
    \lineheight{1}%
    \setlength\tabcolsep{0pt}%
    \put(0,0){\includegraphics[width=\unitlength,page=1]{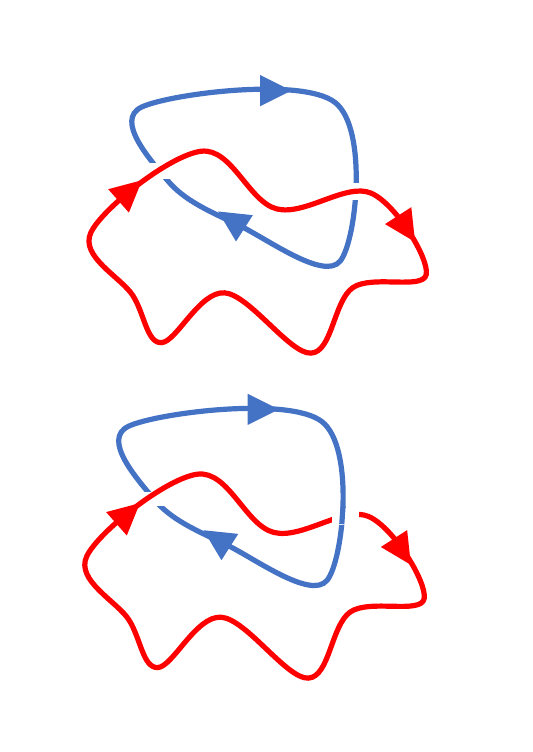}}%
    \put(0.04548804,1.31464518){\color[rgb]{0,0,0}\makebox(0,0)[lt]{\lineheight{1.25}\smash{\begin{tabular}[t]{l}$l(\gamma_1,\gamma_2)=0$\end{tabular}}}}%
    \put(0.00522747,0.9416199){\color[rgb]{0,0,0}\makebox(0,0)[lt]{\lineheight{1.25}\smash{\begin{tabular}[t]{l}$\gamma_1$\end{tabular}}}}%
    \put(0.00234954,0.32881627){\color[rgb]{0,0,0}\makebox(0,0)[lt]{\lineheight{1.25}\smash{\begin{tabular}[t]{l}$\gamma_1$\end{tabular}}}}%
    \put(0.69526203,1.173276){\color[rgb]{0,0,0}\makebox(0,0)[lt]{\lineheight{1.25}\smash{\begin{tabular}[t]{l}$\gamma_2$\end{tabular}}}}%
    \put(0.67013789,0.57577277){\color[rgb]{0,0,0}\makebox(0,0)[lt]{\lineheight{1.25}\smash{\begin{tabular}[t]{l}$\gamma_2$\end{tabular}}}}%
    \put(0.06792214,0.02802022){\color[rgb]{0,0,0}\makebox(0,0)[lt]{\lineheight{1.25}\smash{\begin{tabular}[t]{l}$l(\gamma_1,\gamma_2)=+1$\end{tabular}}}}%
  \end{picture}%
\endgroup%
\label{fig:linkNumberModel}
	}
	\subfloat[]{
	\def\svgscale{0.3}
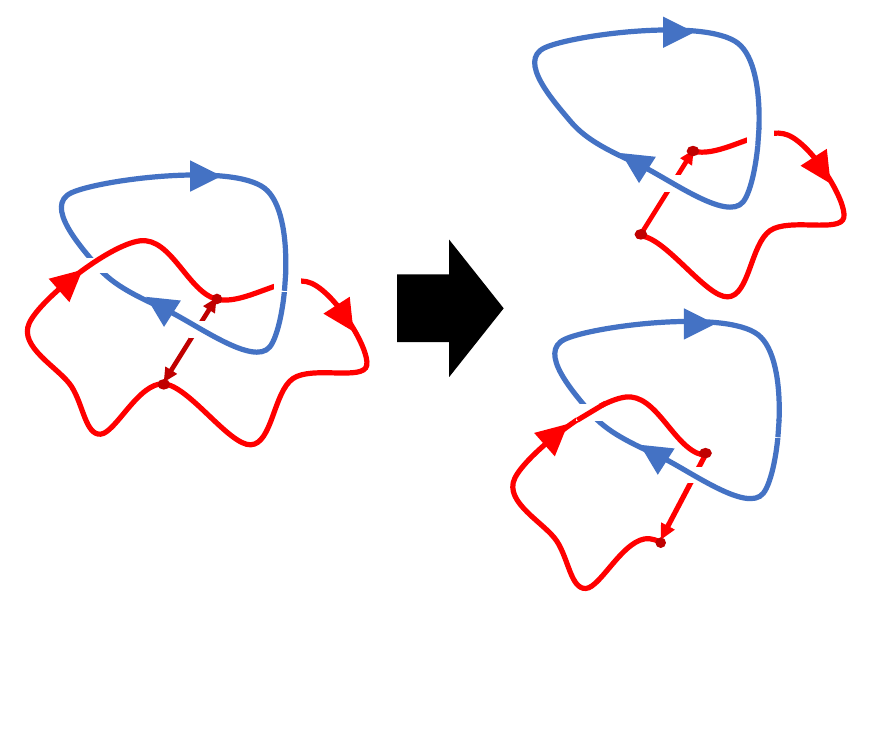\label{fig:curveDecomposition}
	}
	\caption{
	Ensnarled networks and linking numbers:
	\subR{fig:linkNumberModel} The invariant~${l\bra{\gamma_1, \gamma_2}}$ characterizes the topological relation up to a sign, with~${l=0}$ denoting the unlink and~${l=\pm 1}$ trivial links, e.g. the Hopf-link.
	\subR{fig:curveDecomposition} Chords (dark-red) allow for evaluation of~${l\bra{\gamma_1, \gamma_2}}$ via the decomposition into subcurves.
	}\label{fig:modelLink}
\end{figure}

Recall that linking numbers,~${l\in \Z}$, are topological invariants which characterize the linking state (up to sign) of two closed curves, with~${l=0}$ denoting the unlink and~${l=\pm 1, \pm 2, ...}$ denoting linked states where the curves are not separable without breaking at least one of them~\cite{RN405}, see Figure~\ref{fig:linkNumberModel}.
Given two closed, oriented curves  with parametrizations~${\gamma_1,\gamma_2:[0,1) \longrightarrow\R^3}$, their linking number is computable via the Gauss linking integral:
\alg{
	l \bra{\gamma_1, \gamma_2} &= \frac{1}{4\pi}\int_{\gamma_1} \int_{\gamma_2} \bra{d\vc{r}_1\times d\vc{r}_2} \cdot \frac{\vc{r}_1 - \vc{r}_2}{|\vc{r}_1-\vc{r}_2|^3}
\label{eq:linkingIntegral}
}
Note, the Gauss linking integral is a symmetric, bilinear  mapping~\cite{RN405}. Thus a flip in the orientation of one curve~${\gamma_i \longrightarrow -\gamma_i}$ will result in a sign change of the linking number as
\alg{
	l \bra{-\gamma_1, \gamma_2}  =- l \bra{\gamma_1, \gamma_2} =l \bra{\gamma_1, -\gamma_2}
}
One may further rewrite any closed curves as the composition of 'smaller' closed curves~${\gamma_1 = \bigcup_{i=0}^{n} \sigma_{k_i}\gamma_{1k_i}}$ (with canceling chords, see Figure~\ref{fig:curveDecomposition}) with sub-orientations~${\sigma_{k_i}=\pm 1}$ and calculate its linking number as
\alg{
&	l \bra{\gamma_1 =\bigcup_{i=0}^{n} \sigma_{k_i}\gamma_{1k_i}, \gamma_2} \label{eq:linkingNumberDecomp}  \\
&=\sum^{n_1}_{i=1}   \frac{\sigma_{k_i}}{4\pi}\int_{\gamma_{1k_i}} \int_{\gamma_2} \bra{d\vc{r}_{1k_i}\times d\vc{r}_2} \cdot \frac{\vc{r}_{1k_i}- \vc{r}_2}{|\vc{r}_{1k_i}-\vc{r}_2|^3}\nonumber	\\
&=\sum^{n_1}_{i=1} \sigma_{k_i}  l \bra{\gamma_{1k_{i}}, \gamma_2} \nonumber
}

\begin{figure} [t]
	\def\svgscale{0.2}
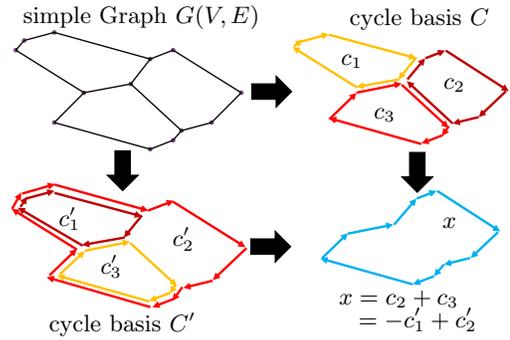
	\caption{
	Constructing arbitrary graph cycles from basis cycle superposition by oriented curve decomposition, as used in the path integrals of linking number calculations, see Figure~\ref{fig:modelLink} and Equation \eqref{eq:cycleLinkDecomp}.
	}\label{fig:cycleBasis}
\end{figure}

Recall that a simple graph~${G}$ is defined as a set of vertices,~${V}$, and edges,~${E}$.
Further, there are two maps~${\alpha, \omega}$ with~${\alpha: E \rightarrow V}$ and~${\omega: E \rightarrow V}$, uniquely defining each edge~${e \in E}$ as a tuple of vertices~${\bra{ \alpha\bra{e},  \omega\bra{e} }}$.
Assuming~${\alpha\bra{e} \neq  \omega\bra{e}}$ one can thereby define the edge's direction.
A path is a sequence of such edges, while a cycle is a path, which starts and ends on the same vertex, traversing any of its edges only once.
Fortunately one only need a certain subset of all possible cycles in a graph, the 'fundamental cycles', to construct the rest.
One may identify these distinct cycles in the following way:
Given a graph with one connected component of~${|V|}$ vertices and~${|E|}$ edges, a subset of~${|V|-1}$ edges is sufficient to connect every vertex and thereby create a spanning tree.
Therefore a set of non-utilized edges remains of cardinality
\g{
	z= |E|-|V|+1 \label{eq:cycles}
}
Adding any of these surplus edges to the tree will result in a fundamental cycle, which can be combined with with other such cycles to account for any possible loop in the graph.
Note that~${z}$ represents the dimension of the graph's cycle space (referred to as 'nullity' or the 'cyclomatic number'~\cite{Whitney1992}), a vector space over~${\Z_2}$ with linear combinations achieved through edgewise XOR operations~\cite{RN229}.
The astute reader may recognize this construction as the first homology of a chain complex composed of only 1-chains \cite{RN435}. 
While this observation suggests an interesting path to generalization, we restrict ourselves in this letter to considering only space curves and network cycles, i.e. 1-chains.
Further, as we are considering spatially embedded graphs, every node, edge and therefore path has a spatial curve representation.
We thus use the decomposition of oriented closed curves into sub-curves for cycles in the spatial graphs considered, see Figure~\ref{fig:cycleBasis}.
Furthermore, one may derive matrices from graphs~${G\bra{V,E,\alpha,\omega}}$ utilizing these topological characteristics, e.g. the 'mesh' matrix typically encountered in electric circuits~\cite{TN_libero_mab214116274},
\g{
	M_{ei}=\begin{cases}
				\pm 1  & \text{ if cycle i contains edge e }\\
				0 & \text{else}
				\end{cases}
				\label{eq:meshMatrix}
}
The sign of~${M_{ei}}$ captures the direction of the cycle with respect to the edge's direction.
Note that~${z}$ equals the rank of the mesh matrix~${\vc{M}}$, and that this operator transforms between the edge-space and cycle-space of a graph.
\\
Let us consider cycle bases for two graphs as~${C_1 = \bro{c_{11}, ... , c_{1z_1}}}$ and~${C_2 = \bro{c_{21}, ... , c_{2z_2}}}$ and a polygonal representation via the respective graph's edges (piece-wise straight lines connecting the vertices).
As previously shown, we may decompose any linking number calculation of arbitrary cycles~${c_1, c_2}$ in a graph pair as
\alg{
	l \bra{c_1, c_2} &=\sum^{z_1, z_2}_{i, j} \sigma_{1i} \sigma_{2j}  l \bra{c_{1i}, c_{2j} } \label{eq:cycleLinkDecomp}
}
\begin{figure*}[t]
	\hspace{-2.25cm}
	\begin{minipage}{4cm}
	\hspace{1cm}
		\includegraphics[scale=0.25]{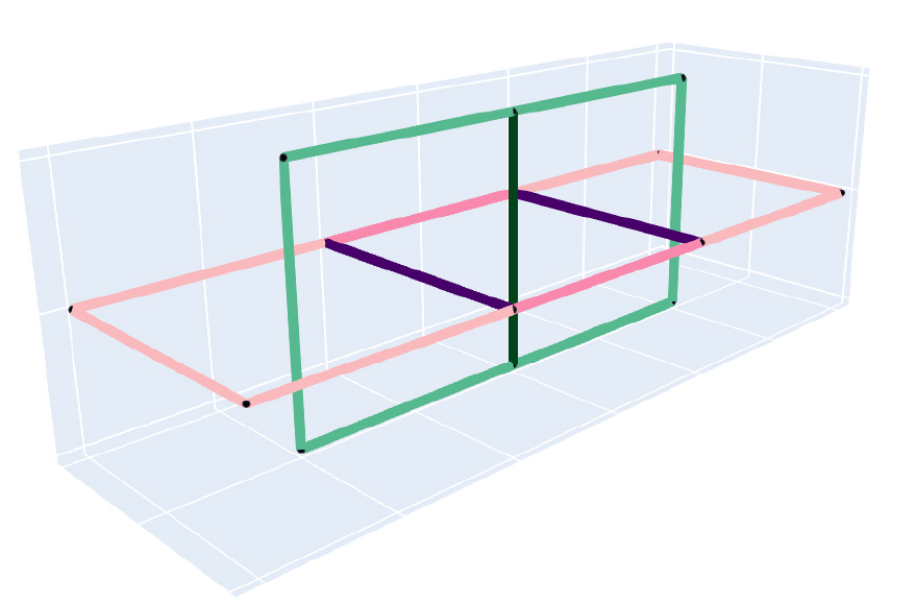}\\
		\subfloat[]{
			\includegraphics[scale=1.]{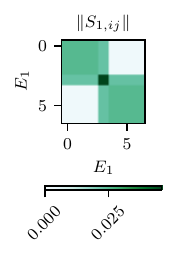}\hspace{-0.5cm}
			\includegraphics[scale=1.]{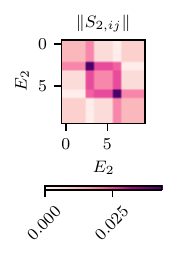}
	}
	\end{minipage}
	\hspace{1.75cm}
	\begin{minipage}{4cm}
	\hspace{1cm}
		\includegraphics[scale=0.25]{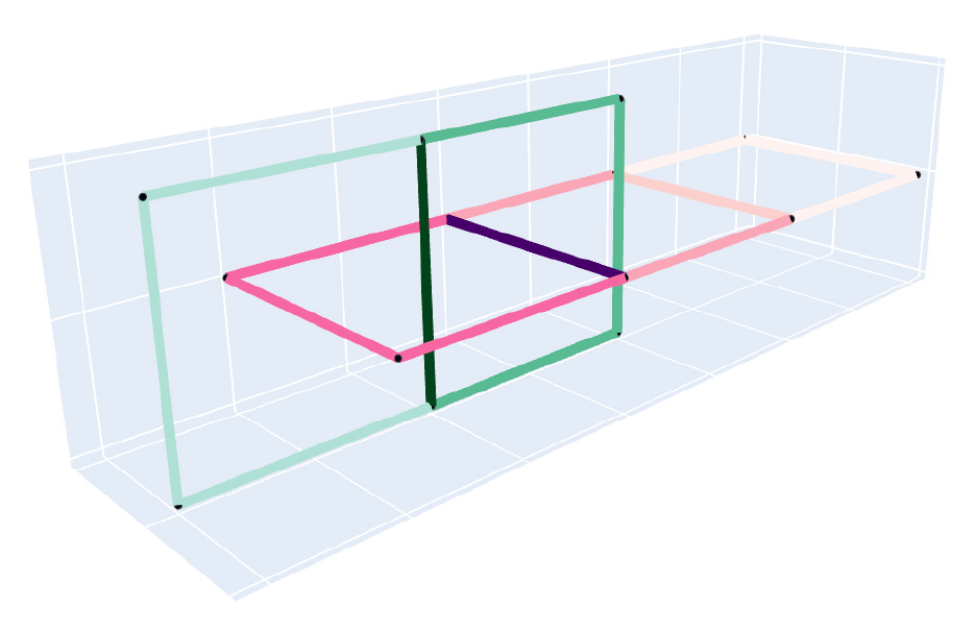}
		\subfloat[]{
			\includegraphics[scale=1.]{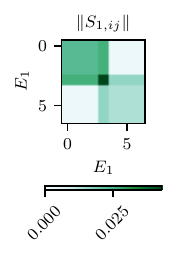}\hspace{-0.5cm}
			\includegraphics[scale=1.]{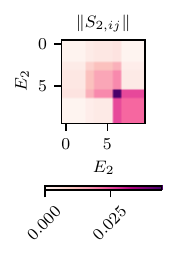}
	}
	\end{minipage}
	\hspace{1.75cm}
	\begin{minipage}{4cm}
	\hspace{1cm}
		\includegraphics[scale=0.25]{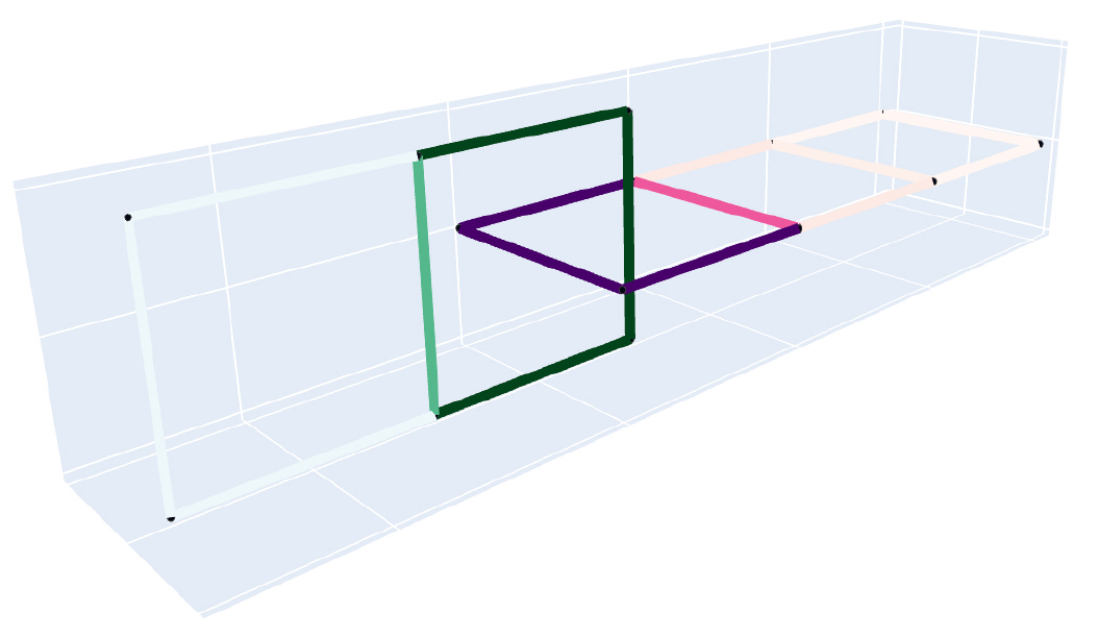}
		\subfloat[]{
			\includegraphics[scale=1.]{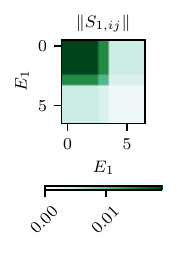}\hspace{-0.5cm}
			\includegraphics[scale=1.]{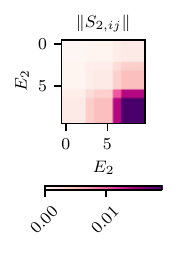}
		}
	\end{minipage}
	\caption{
	Fully linked meshes unlink via axial translation.
	Top row: Visualization of simple meshes, color coded according to the diagonals of the corresponding priority operator $\vc{S}_i$.
	Bottom row: Heatmaps depicting the absolute values of priority operators ${\vc{S}_i}$ for each mesh. 
	}\label{fig:DualLadder}
\end{figure*}

Or in terms of a linking number operator~${\vc{L}}$:
\alg{
l \bra{c_1, c_2} &= \vc{\sigma}_1^T\cdot \vc{L} \cdot \vc{\sigma}_2 \label{eq:vecLK}\\
\text{ with } L_{ij} &= l \bra{c_{1i}, c_{2j} } \text{ , } \vc{L}\in \Z^{z_1\times z_2}, \vc{\sigma}_i \in \Z^{z_i}\nonumber
}
Naturally, the linking numbers in this bilinear decomposition are invariant with respect to a change of the cycle basis, e.g. to~${C_1'}$ and~${C_2'}$,
\alg{
\vc{\sigma}_1^T\cdot \vc{L} \cdot \vc{\sigma}_2 = \vc{\sigma}_1^{'T}\cdot \vc{L}^{'} \cdot \vc{\sigma}_2^{'}
}
demonstrating the independence to the choice of bases per graph.
Next, we construct the invariant operator~$\vc{\Lambda}$ which determines the importance of individual edges.
Subsequently, we factorize~${\vc{L}}$ using the mesh matrices~${\vc{M}_i}$ of the respective networks (corresponding to the current basis),
\alg{
	\vc{L} =\vc{M}_{1}^T \cdot \vc{\Lambda} \cdot  \vc{M}_{2}
}
The matrix~${\vc{\Lambda} \in \R^{E1\times E_2}}$ is constructed via generalized inverses, as the mesh matrices do not have full rank:
\alg{
	\vc{\Lambda} = \brc{\vc{M}_1^{\dagger} +\vc{X}_1\cdot \vc{P}_1}^T \cdot \vc{L} \cdot \brc{\vc{M}_2^{\dagger}+\vc{X}_2\cdot\vc{P}_2}
}
with arbitrary factors~${\vc{X}_i \in \R^{z_i \times E_i}}$ and orthogonal projectors
\alg{
\vc{P}_1 = \brc{ \vc{I} - \vc{M}_1\vc{M}_1^{\dagger}} \text{ and } \vc{P}_2 = \brc{ \vc{I} - \vc{M}_2\vc{M}_2^{\dagger}}
}
These projectors signal a gauge freedom for~${\vc{\Lambda}}$ as they satisfy~${\vc{P}_i\vc{M}_i=0}$.
Here we choose the least square solution as the canonical form:
\alg{
	\vc{\Lambda}_0 \equiv \vc{M}_1^{\dagger , T}  \cdot \vc{L} \cdot\vc{M}_2^{\dagger}
}
We may then rewrite the linking number of two arbitrary cycles in a graph pair as
\alg{
l \bra{c_1, c_2} &= \vc{\sigma}_1^T\cdot \vc{M}_{1}^T \cdot \vc{\Lambda}_0 \cdot  \vc{M}_{2}\cdot \vc{\sigma}_2
}
where we know the edge representation of the cycles to be
\alg{
	\vc{p}_1^T =\vc{\sigma}_1^T\cdot \vc{M}_{1}^T \text{ and } \vc{p}_2 = \vc{M}_{2}\cdot \vc{\sigma}_2
	\label{eq:meshMatrixProjection}
}
with~${\vc{p}_i\in K^{E_i}}$, where~$K=\bro{-1, 0, 1}$.
Thus we may write the linking number of any two cycles in the ensnarled system in edge-space form as
\alg{
l \bra{c_1, c_2} &=\vc{p}_1^T\cdot \vc{\Lambda}_0 \cdot \vc{p}_2
}
\newline

\section{Results}
\label{sec:results}
\begin{figure}[t]
\hspace{-0.45cm}
\subfloat[]{
\begin{minipage}{4.35cm}
\includegraphics[scale=.65]{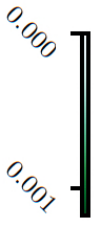}
\hspace{-0.35cm}
	\def\svgscale{0.325}
\begingroup%
  \makeatletter%
  \providecommand\color[2][]{%
    \errmessage{(Inkscape) Color is used for the text in Inkscape, but the package 'color.sty' is not loaded}%
    \renewcommand\color[2][]{}%
  }%
  \providecommand\transparent[1]{%
    \errmessage{(Inkscape) Transparency is used (non-zero) for the text in Inkscape, but the package 'transparent.sty' is not loaded}%
    \renewcommand\transparent[1]{}%
  }%
  \providecommand\rotatebox[2]{#2}%
  \newcommand*\fsize{\dimexpr\f@size pt\relax}%
  \newcommand*\lineheight[1]{\fontsize{\fsize}{#1\fsize}\selectfont}%
  \ifx\svgwidth\undefined%
    \setlength{\unitlength}{283.21101379bp}%
    \ifx\svgscale\undefined%
      \relax%
    \else%
      \setlength{\unitlength}{\unitlength * \real{\svgscale}}%
    \fi%
  \else%
    \setlength{\unitlength}{\svgwidth}%
  \fi%
  \global\let\svgwidth\undefined%
  \global\let\svgscale\undefined%
  \makeatother%
  \begin{picture}(1,0.92176869)%
    \lineheight{1}%
    \setlength\tabcolsep{0pt}%
    \put(0,0){\includegraphics[width=\unitlength,page=1]{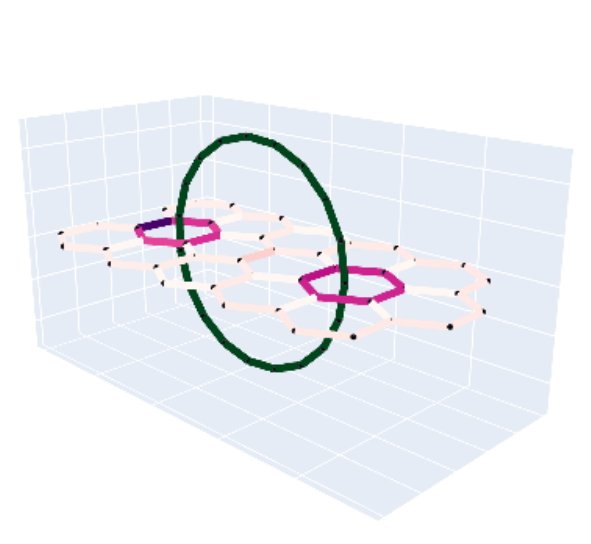}}%
    \put(0.08658893,0.82570455){\color[rgb]{0,0,0}\makebox(0,0)[lt]{\lineheight{1.25}\smash{\begin{tabular}[t]{l}1. Iteration\end{tabular}}}}%
    \put(0,0){\includegraphics[width=\unitlength,page=2]{hopfedHexagon_0_annotated.pdf}}%
  \end{picture}%
\endgroup%
\\
	\includegraphics[scale=.65]{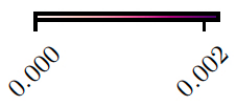}
\end{minipage}
}
\hspace{-0.45cm}
\subfloat[]{
\begin{minipage}{4.45cm}
\includegraphics[scale=.65]{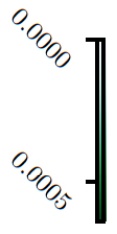}
\hspace{-0.35cm}
	\def\svgscale{0.325}
\begingroup%
  \makeatletter%
  \providecommand\color[2][]{%
    \errmessage{(Inkscape) Color is used for the text in Inkscape, but the package 'color.sty' is not loaded}%
    \renewcommand\color[2][]{}%
  }%
  \providecommand\transparent[1]{%
    \errmessage{(Inkscape) Transparency is used (non-zero) for the text in Inkscape, but the package 'transparent.sty' is not loaded}%
    \renewcommand\transparent[1]{}%
  }%
  \providecommand\rotatebox[2]{#2}%
  \newcommand*\fsize{\dimexpr\f@size pt\relax}%
  \newcommand*\lineheight[1]{\fontsize{\fsize}{#1\fsize}\selectfont}%
  \ifx\svgwidth\undefined%
    \setlength{\unitlength}{278.97196198bp}%
    \ifx\svgscale\undefined%
      \relax%
    \else%
      \setlength{\unitlength}{\unitlength * \real{\svgscale}}%
    \fi%
  \else%
    \setlength{\unitlength}{\svgwidth}%
  \fi%
  \global\let\svgwidth\undefined%
  \global\let\svgscale\undefined%
  \makeatother%
  \begin{picture}(1,0.88190955)%
    \lineheight{1}%
    \setlength\tabcolsep{0pt}%
    \put(0,0){\includegraphics[width=\unitlength,page=1]{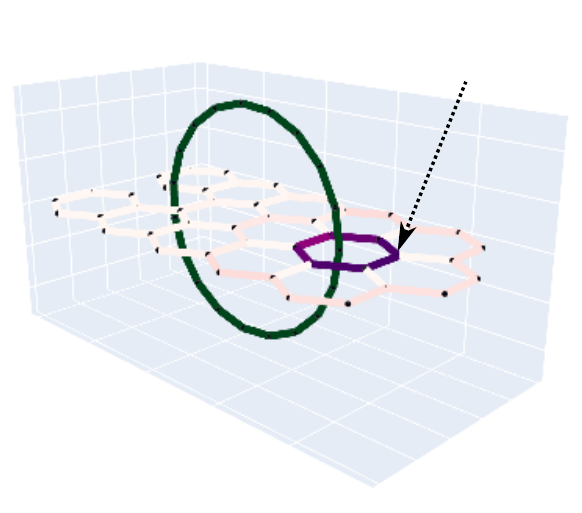}}%
    \put(0.37185941,0.83422111){\color[rgb]{0,0,0}\makebox(0,0)[lt]{\lineheight{1.25}\smash{\begin{tabular}[t]{l}2. Iteration\end{tabular}}}}%
    \put(0,0){\includegraphics[width=\unitlength,page=2]{hopfedHexagon_1_annotated.pdf}}%
  \end{picture}%
\endgroup%
\\
	\includegraphics[scale=.65]{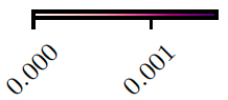}
\end{minipage}
}
\\
\hspace{-0.45cm}
\subfloat[]{
\begin{minipage}{4.45cm}
\includegraphics[scale=.65]{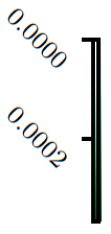}
\hspace{-0.35cm}
	\def\svgscale{0.325}
\begingroup%
  \makeatletter%
  \providecommand\color[2][]{%
    \errmessage{(Inkscape) Color is used for the text in Inkscape, but the package 'color.sty' is not loaded}%
    \renewcommand\color[2][]{}%
  }%
  \providecommand\transparent[1]{%
    \errmessage{(Inkscape) Transparency is used (non-zero) for the text in Inkscape, but the package 'transparent.sty' is not loaded}%
    \renewcommand\transparent[1]{}%
  }%
  \providecommand\rotatebox[2]{#2}%
  \newcommand*\fsize{\dimexpr\f@size pt\relax}%
  \newcommand*\lineheight[1]{\fontsize{\fsize}{#1\fsize}\selectfont}%
  \ifx\svgwidth\undefined%
    \setlength{\unitlength}{281.42328644bp}%
    \ifx\svgscale\undefined%
      \relax%
    \else%
      \setlength{\unitlength}{\unitlength * \real{\svgscale}}%
    \fi%
  \else%
    \setlength{\unitlength}{\svgwidth}%
  \fi%
  \global\let\svgwidth\undefined%
  \global\let\svgscale\undefined%
  \makeatother%
  \begin{picture}(1,0.90394094)%
    \lineheight{1}%
    \setlength\tabcolsep{0pt}%
    \put(0,0){\includegraphics[width=\unitlength,page=1]{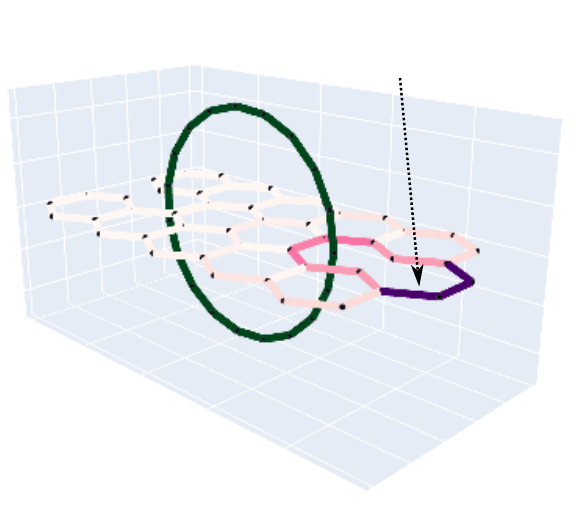}}%
    \put(0.27093598,0.84611829){\color[rgb]{0,0,0}\makebox(0,0)[lt]{\lineheight{1.25}\smash{\begin{tabular}[t]{l}3. Iteration\end{tabular}}}}%
    \put(0,0){\includegraphics[width=\unitlength,page=2]{hopfedHexagon_2_annotated.pdf}}%
  \end{picture}%
\endgroup%
\\
	\includegraphics[scale=.65]{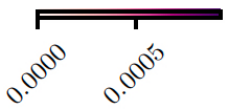}
\end{minipage}
}
\hspace{-0.45cm}
\subfloat[]{
	\begin{minipage}{4.35cm}
\includegraphics[scale=.65]{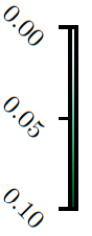}
\hspace{-0.4cm}
	\def\svgscale{0.325}
\begingroup%
  \makeatletter%
  \providecommand\color[2][]{%
    \errmessage{(Inkscape) Color is used for the text in Inkscape, but the package 'color.sty' is not loaded}%
    \renewcommand\color[2][]{}%
  }%
  \providecommand\transparent[1]{%
    \errmessage{(Inkscape) Transparency is used (non-zero) for the text in Inkscape, but the package 'transparent.sty' is not loaded}%
    \renewcommand\transparent[1]{}%
  }%
  \providecommand\rotatebox[2]{#2}%
  \newcommand*\fsize{\dimexpr\f@size pt\relax}%
  \newcommand*\lineheight[1]{\fontsize{\fsize}{#1\fsize}\selectfont}%
  \ifx\svgwidth\undefined%
    \setlength{\unitlength}{287.66173553bp}%
    \ifx\svgscale\undefined%
      \relax%
    \else%
      \setlength{\unitlength}{\unitlength * \real{\svgscale}}%
    \fi%
  \else%
    \setlength{\unitlength}{\svgwidth}%
  \fi%
  \global\let\svgwidth\undefined%
  \global\let\svgscale\undefined%
  \makeatother%
  \begin{picture}(1,0.85060239)%
    \lineheight{1}%
    \setlength\tabcolsep{0pt}%
    \put(0,0){\includegraphics[width=\unitlength,page=1]{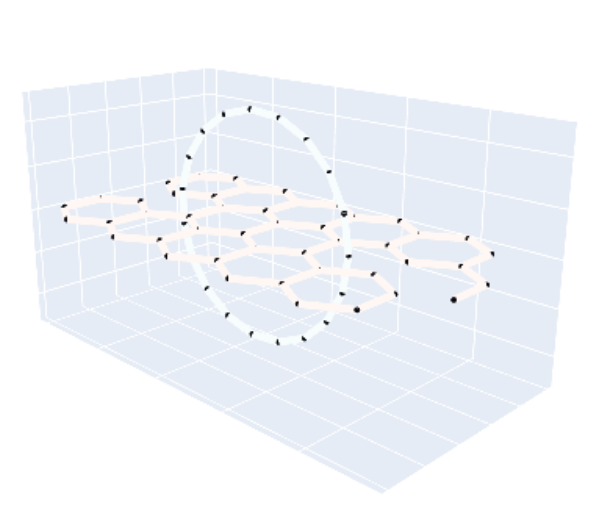}}%
    \put(0.03574939,0.86462649){\color[rgb]{0,0,0}\makebox(0,0)[lt]{\lineheight{1.25}\smash{\begin{tabular}[t]{l}4. Iteration\end{tabular}}}}%
    \put(0,0){\includegraphics[width=\unitlength,page=2]{hopfedHexagon_3_annotated.pdf}}%
    \put(0.02891568,0.77108431){\color[rgb]{0,0,0}\makebox(0,0)[lt]{\lineheight{1.25}\smash{\begin{tabular}[t]{l}No link detected\end{tabular}}}}%
  \end{picture}%
\endgroup%
\\
	\includegraphics[scale=.65]{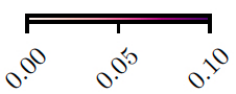}
\end{minipage}
}
	\caption{
	Greedy edge cutting algorithm: Example of a single ring linked to a hexagonal mesh, unlinked in 4 iterations. 
	Displayed is the momentary optimal cut, based on highest edge priority.
	}\label{fig:cuttingEdge}
\end{figure}
The meaning of~${\vc{\Lambda}_0}$ becomes more apparent if we consider the symmetric constructions of the positive definite operators~${\vc{S}_1=\vc{\Lambda}_0\vc{\Lambda}_0^T \in \R^{E_1\times E_1}}$ and~${\vc{S}_2=\vc{\Lambda}_0^T\vc{\Lambda}_0\in \R^{E_2\times E_2}}$.
We focus on these operators in particular as they allow the evaluation of the impact of edges on non-zero linking numbers.
\\
As an illustration of the concept, consider a dual ladder graph system, as shown in Figure~\ref{fig:DualLadder}.
We display the changes in~${\vc{S}_i}$ while separating the graphs in discrete translational steps.
We refer to the diagonals of~${\vc{S}_i}$ as the \textit{priority} of edges, see appendix A for further detail.
The plots show the respective edge priorities, the visualization of the respective diagonals of~${\vc{S}_i}$.
Note that priorities are at their maximum for the most central ladder legs, corresponding to the edges which must be traversed by any linked cycle.
As the graphs are separated one observes a shift of edge priorities in both graphs as different edges become crucial connections through which any linked cycle must pass.
A complete spatial separation of the networks naturally results in the~${\vc{S}_i}$ becoming null operators as no link remains.
One may therefore characterize any ensnarled system by a hierarchical ordering of edge priorities, indicating the crucial topological links in both graphs.
Such a scheme may be used to determine the optimal cut between two meshes, i.e. \textit{how many and which edges}~${\varepsilon}$ one has to remove from either graph such that the entire system becomes unlinked.
\newline
We propose the following simple algorithm to identify such a cut set:
Given two graphs~${G_1}$ and~${G_2}$, set one as a reference graph~${G_{ref}}$ and the other as a target graph~${G_{target}}$.
Then compute the linkage matrix~${\vc{L}}$ as described before and the edge space operators~${\vc{\Lambda}_0}$,~${\vc{S}_{ref}}$ and~${\vc{S}_{target}}$.
Next, identify the maximum priority on the diagonal of~${\vc{S}_{target}}$ and remove the corresponding edge from~${G_{target}}$.
Iterate on the linking number evaluation and the edge removal until the linking number calculation generates null-matrices, indicating the unlink of the original meshes~${G_1}$,~${G_2}$.
Doing so yield a list of edges in the target graph representing a cut set of cardinality~${\varepsilon_{target}}$.
\newline
Despite the fact that this is a greedy algorithm, we found it to reliably identify efficient cuts between the meshes, independent of their particular embedding and graph topology, as for example demonstrated for the Hopf-linked hexagonal grid in Figure~\ref{fig:cuttingEdge}.
Note though that we only consider the edges of highest priority during each iteration, which may lead to cuts not representative of the global optimum.
Such inaccuracies occur when meshes have links close to their periphery, leading to high edge priorities in this section which may obscure better cuts.
For example, in Figure~\ref{fig:cuttingEdge}, there is an equally efficient cut set found by removing the edges from the hexagonal grid passing the rings cross-section, yet edges on the periphery of the hexagon have higher priorities.
Additional deviations may occur as the number of degree $k=2$ vertices is varied along paths in the network (which may anytime be done by for example by adding new vertices in a straight line between two already existing ones forming an edge), appendix A.
On that line we propose that coarse-graining steps should be taken into account, removing as many k=2 vertices as possible, leaving the linkage of the system intact, as to reduce the complexity of the problem beforehand. \\
Further, this algorithm also suggests an associated metric for ensnarled systems.
Let us refer to the cardinality of cuts for any mesh~${G_i}$ as~${\varepsilon_i}$ and compare it to the cyclomatic number~${z_i}$.
We accordingly define the relative cut number as
\alg{
\rho_i \equiv \frac{\varepsilon_i}{z_i}
}
accounting for how much of a network's cycle space has to be taken down before the unlinking event, i.e. if~${\rho_i=1}$ every single cycle in a mesh has to be eliminated.
The trivial case of~${z_i=0}$ simply implies that at least one of the networks is a tree, and therefore not topologically linked by default.
We are not only able to identify which edges to remove to break the topological linkage, we are also able to account for asymmetries and redundancies.
This metric classifies networks according to their basis utilization, indicating whether a network has a bloated cycle space which does not significantly contribute to the linkage of the system.
\begin{figure}[t]
	\begin{minipage}{2.5cm}
		\centering
		\includegraphics[scale=0.11]{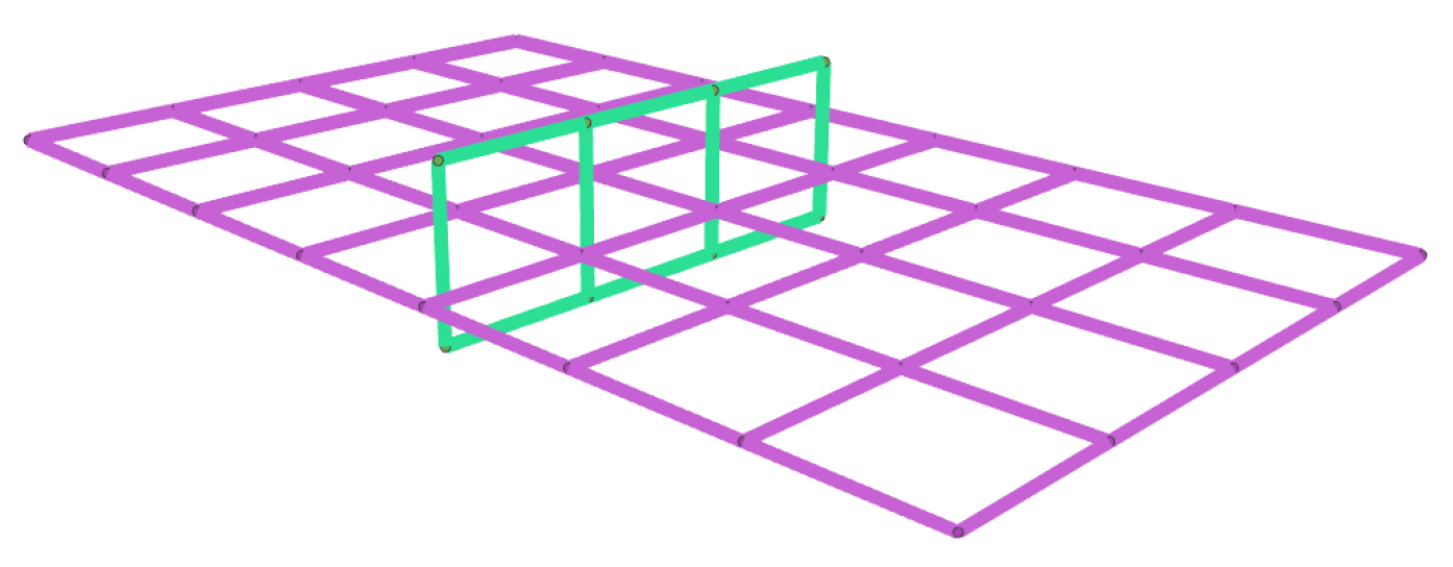}
		meshV1
	\end{minipage}
	\begin{minipage}{2.5cm}
		\centering
		\includegraphics[scale=0.15]{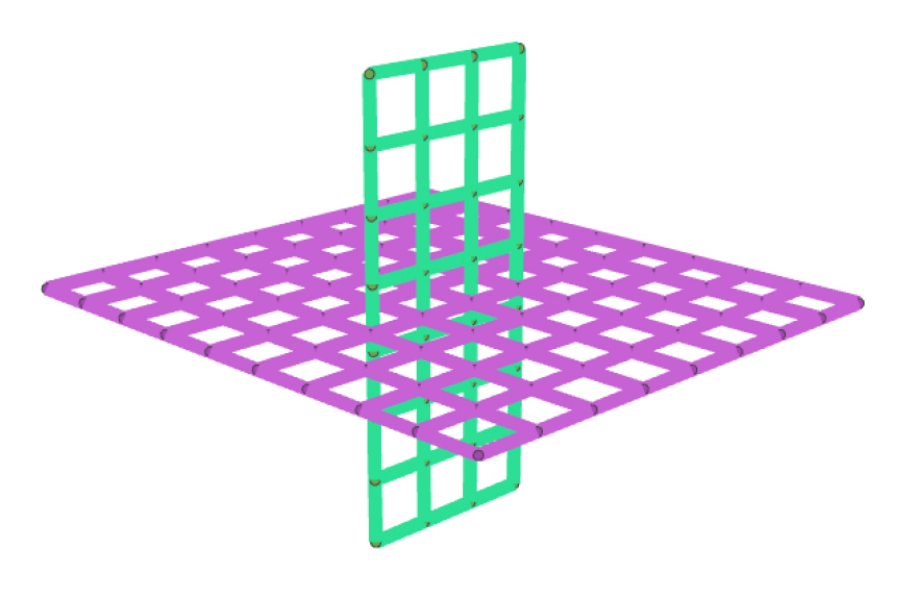}\\
		meshV2
	\end{minipage}
	\begin{minipage}{2.5cm}
		\centering
		\includegraphics[scale=0.125]{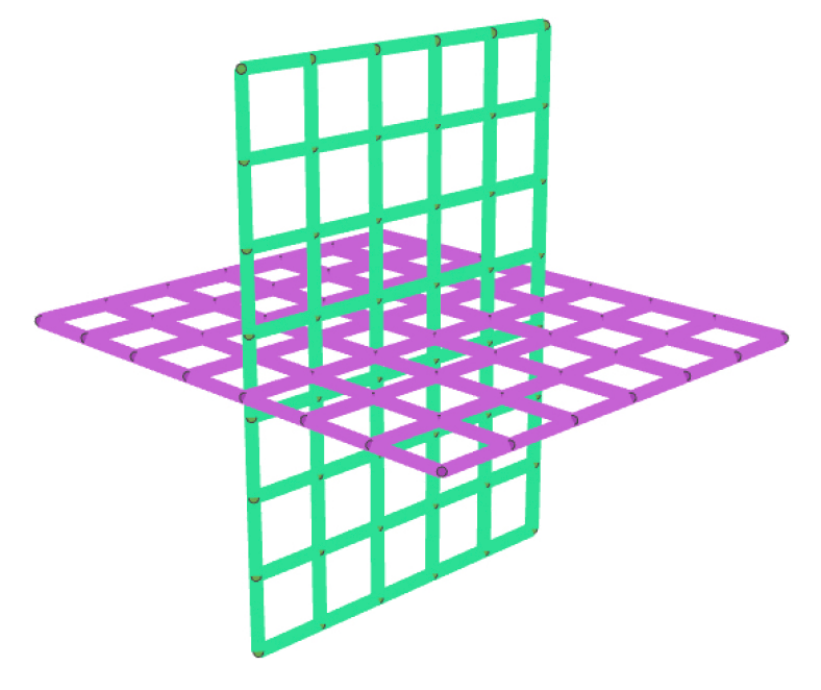}
		meshV3
	\end{minipage}
\\
	\begin{minipage}{5cm}
		\includegraphics[scale=1.]{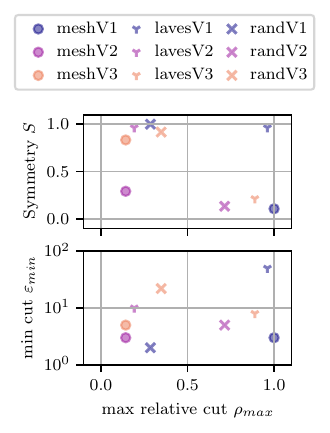}
	\end{minipage}
\hspace{0.25cm}
	\begin{minipage}{3cm}
		\includegraphics[scale=0.15]{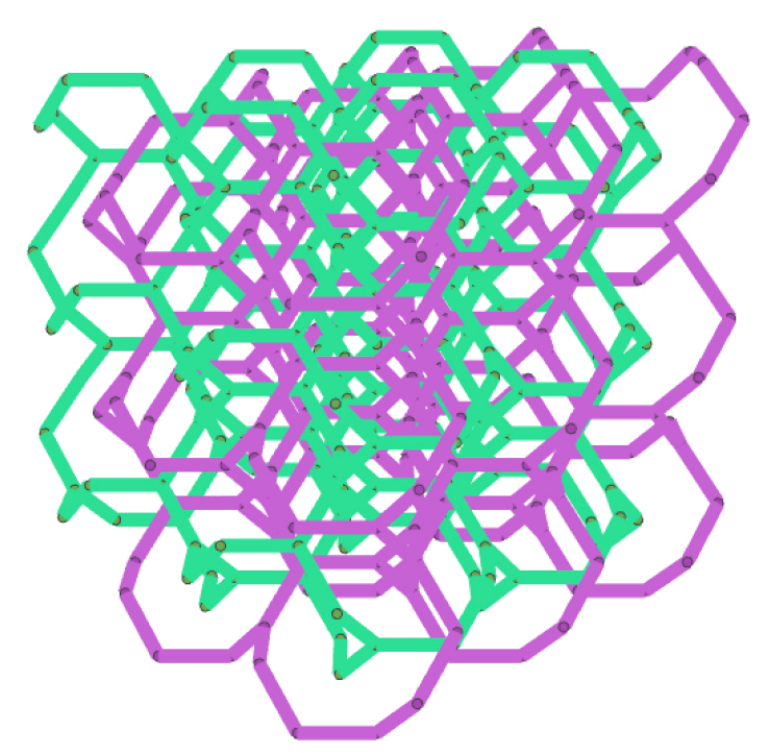}\\
		lavesV1\\
		\includegraphics[scale=0.15]{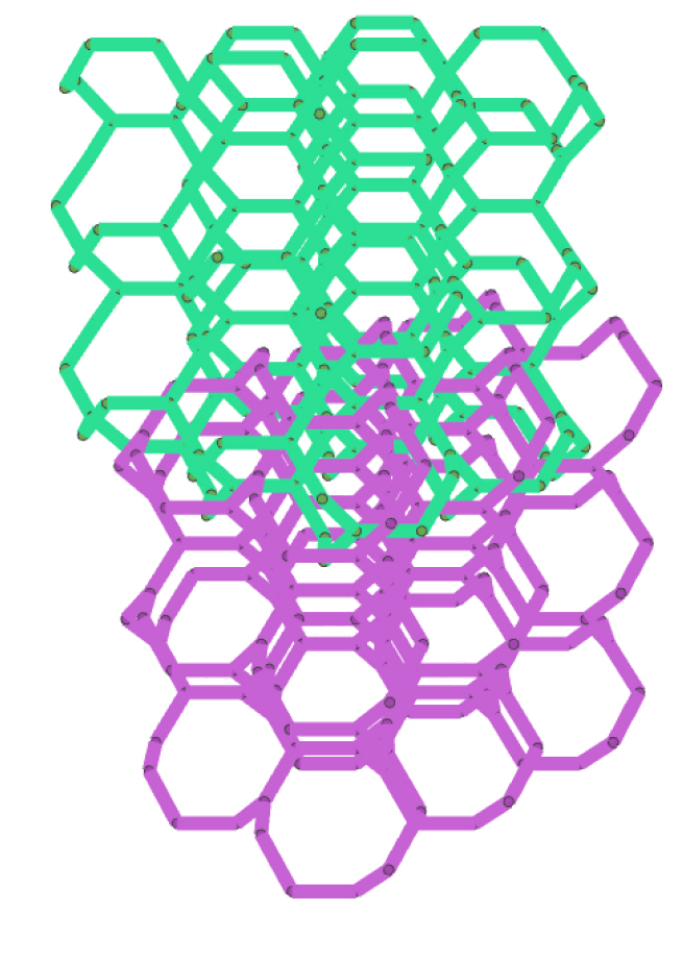}\\
		lavesV2\\
		\includegraphics[scale=0.15]{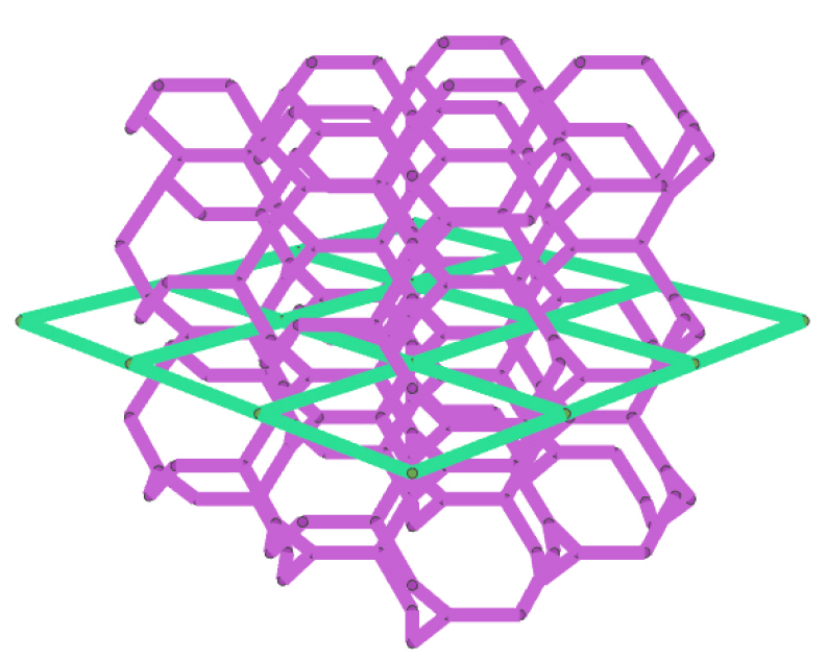}\\
		lavesV3\\
	\end{minipage}\\
	\begin{minipage}{2.5cm}
		\centering
		\includegraphics[scale=0.2]{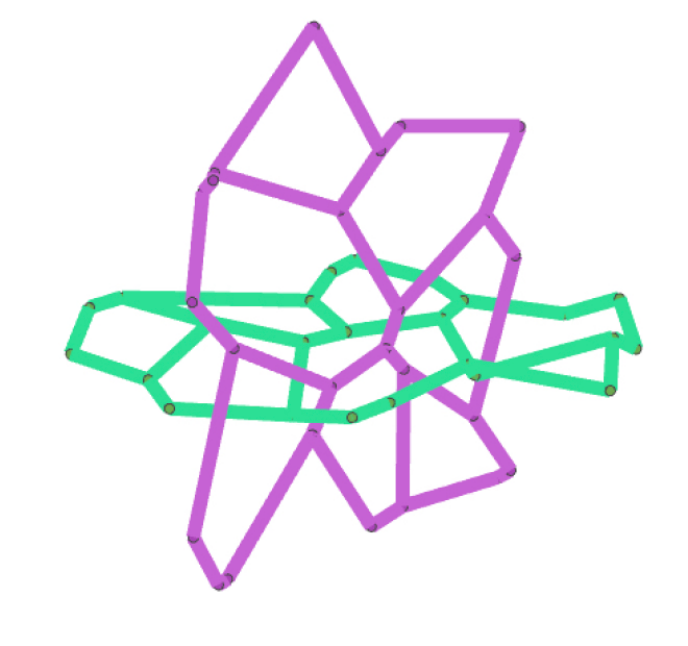}\\
		randV1
	\end{minipage}
	\begin{minipage}{2.5cm}
		\centering
		\includegraphics[scale=0.2]{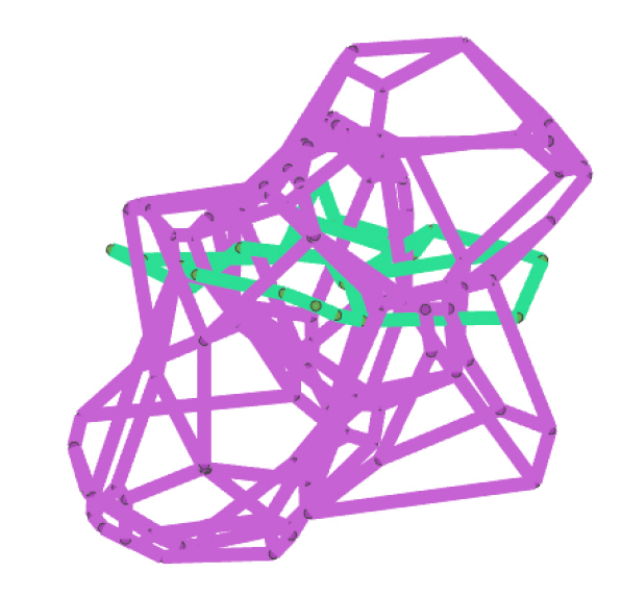}
		randV2
	\end{minipage}
	\begin{minipage}{2.5cm}
		\centering
		\includegraphics[scale=0.2]{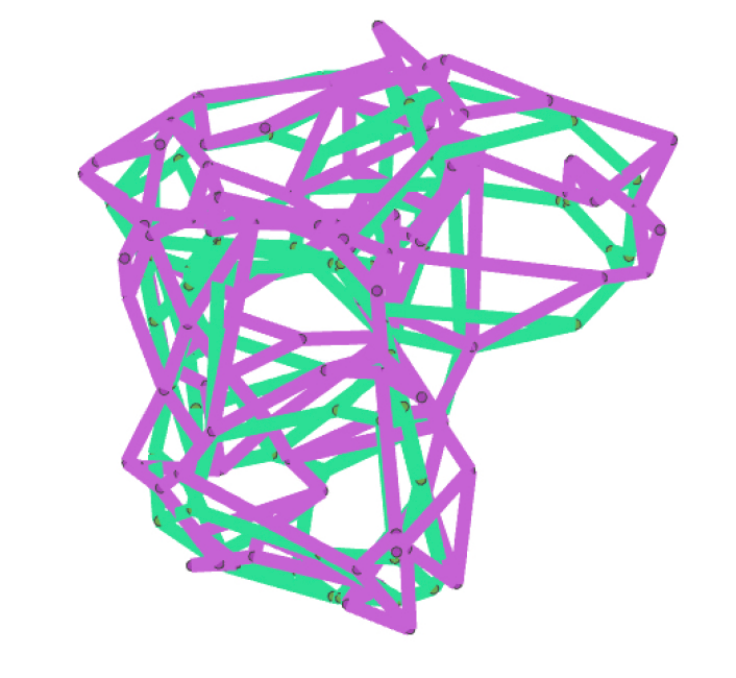}
		randV3
	\end{minipage}
		\caption{
		Classification of entangled systems: Maximum relative cut number~${\rho_{max}}$ \eqref{eq:maxRho} and linkage asymmetry~$S$ \eqref{eq:RatioRho} present a system's redundancy and asymmetry, while the mutual cut coefficient~${\varepsilon_{min}}$ \eqref{eq:MinCut} being the minimal number of cuts between the two systems.
	}\label{fig:classification}
\end{figure}

As one typically has no preference for either of the two meshes involved we utilize the following order parameters:
\alg{
\rho_{max} = max\bra{\rho_1, \rho_2} \label{eq:maxRho}\\
S= \frac{min\bra{\rho_1, \rho_2}}{\rho_{max}} \label{eq:RatioRho}
}
These two parameters effectively allow us to distinguish different ensnarled mesh systems on the basis of their effective cycle space utilization as well as their linkage symmetry.
These tools, however miss size-dependent effects.
To address this, we define the minimal  cut coefficient,
\alg{
 \varepsilon_{min} = min\bra{\varepsilon_1,\varepsilon_2}\label{eq:MinCut}
 }
which accounts for the actual number of cuts one has to perform before the system splits.
This order parameter is thus sensitive the size of the ensnarled system.
We highlight the strength of this method in Figure~\ref{fig:classification}, where we survey a small set of ensnarled archetypes, i.e. finite spatially embedded graphs (either planar or non-planar) with different kinds of cycle space utilization.
Note that our metric is sensitive to translations of highly symmetrical lattices ('Laves' examples) as well as to random spatial structures ('rand' examples), while being able to distinguish networks of equal cut numbers in terms of cycle space utilization and symmetry ('mesh' examples).

\section{Discussion}
\label{sec:discussion}
We have demonstrated a new classification tool for topologically linked spatial networks, by combining conventional methodologies of graph topology, mesh matrices and linking number integrals.
This allows for effective surveys of ensnarled networks on the basis of Hopf link identification and provides the means to classify different redundancies and symmetries in such meshes.
Future studies on this matter certainly will address the  conceptualization of algorithms that would provide us consistent generation of ensnarled spatial networks, e.g. based on reversed engineering of such structures given arbitrary $\vc{\Lambda}_0$ and spatial constraints.
Note further, that the approach is unable to identify higher order links such as Brunnian entanglements~\cite{RN402}.
Possibly, one could extend the method to include these cases by utilizing Milnor invariants~\cite{RN371, Milnor1957} or similar extending the metric to more than two networks at a time. 
Nevertheless we would point to applicability of this approach to vessel systems in developmental and pathological biology, as the liver lobule presents a model to be more thoroughly tested, e.g. for zonation effects, see appendix B. 
Naturally this could also be deployed with regard to other organs such as kidney, pancreas and the lymphatic system or optimized transport networks more generally, see appendix C.
Furthermore, as connectome mapping matures, there is mounting evidence for spatial intertwining among neural subcircuits, presenting another intriguing potential application \cite{RN434}.
\\
We therefore envision this approach to enable a new kind of spatial network analysis, by taking the topological consequences of embedding into account. 
We would like to note that the methods and algorithms deployed in this study have an open-source Python implementation available for further testing and exploration of ensnarled networks: SnarlyPy \cite{snarlpy}.
\begin{acknowledgements}
We would like to dearly thank the members of the Zerial-lab (MPI-CBG) for providing us with experimental data and expertise to analyze liver lobule structures. 
We would further like to thank the members of the Modes- and Zechner-lab (CSBD) for feedback and criticism during the creation of the manuscript.
In particular we would like to thank Szabolcs Horvat and Carlos Duque for their advice during the research phase.
\end{acknowledgements}

\bibliography{./lib/references} 

\end{document}


\title{Ensnarled: On the topological linkage of spatially embedded network pairs}
\author{Felix Kramer}
\email[]{felixuwekramer@proton.me}
\affiliation{Max Planck Institute for Molecular Cell Biology and Genetics (MPI-CBG),  Dresden 01307, Germany}
\affiliation{Center for Systems Biology Dresden (CSBD),  Dresden 01307, Germany}
\author{Carl D. Modes} 
\email[]{modes@mpi-cbg.de}
\affiliation{Max Planck Institute for Molecular Cell Biology and Genetics (MPI-CBG), Dresden 01307, Germany}
\affiliation{Center for Systems Biology Dresden (CSBD),  Dresden 01307, Germany}
\affiliation{Cluster of Excellence, Physics of Life, TU  Dresden, Dresden 01307, Germany}
\date{\today}
\maketitle

\appendix
\section{From circuits to linkages}
\label{sec:appendixD}
This appendix is intended to detail the derivation of the edge priority operator $\vc{\Lambda}_0$.
First let us re-introduce the mesh matrix $\vc{M}$, which was defined in the main article, and is commonly used in linear circuit theory~\cite{TN_libero_mab214116274}.
The mesh matrix $\vc{M}$ (also fundamental loop -, circuit -, tie set matrix) indicates which edges are involved in which of cycle in a graph.
It also provides us with the possibility to project any vectors of the cycle space on to the edge space as
\alg{
	\vc{M} \cdot \vc{\sigma} = \vc{p}
}
In terms of circuit theory one interprets $\vc{\sigma}$ as a vector of mesh currents and $\vc{p}$ the vector of edge currents.
The mesh impedance for a network with homogeneous edge resistance (equal to one) is further calculated as
\alg{
	\vc{Z} = \vc{M}^T \cdot \vc{M}
}
Now, let us turn toward the interpretation of $\vc{\Lambda_0}^T\vc{\Lambda_0}$ and $\vc{\Lambda_0}\vc{\Lambda_0}^T$ as we calculate these as
\alg{
\vc{\Lambda_0}\vc{\Lambda_0}^T &= \vc{M}_1^{\dagger , T}  \cdot \vc{L} \cdot\vc{M}_2^{\dagger}\vc{M}_2^{\dagger , T}  \cdot \vc{L}^T \cdot\vc{M}_1^{\dagger}\\
	\vc{\Lambda_0}^T\vc{\Lambda_0} &=\vc{M}_2^{\dagger , T}  \cdot \vc{L}^T \cdot\vc{M}_1^{\dagger}\vc{M}_1^{\dagger , T}  \cdot \vc{L} \cdot\vc{M}_2^{\dagger}
}
Now as the impedance matrices $\vc{Z}_i=\vc{M}_i^{\dagger}\vc{M}_i^{\dagger , T} $ are actually invertable we may rewrite this as
\alg{
\vc{S}_1 &= \vc{M}_1^{\dagger , T}  \cdot \vc{L} \cdot\vc{Z}_2^{-1}  \cdot \vc{L}^T \cdot\vc{M}_1^{\dagger}\\
	\vc{S}_2 &=\vc{M}_2^{\dagger , T}  \cdot \vc{L}^T \cdot \vc{Z}_1^{-1}  \cdot \vc{L} \cdot\vc{M}_2^{\dagger}
}
As the networks (and their loops) get larger in terms of edges per path, one can readily see that the effective mesh conductivity $\vc{Z}_i^{-1}$ decreases.
Remember, that $\vc{L}$ denotes the linkage matrix of the respective fundamental loop sets.
In comparison with electric circuits one may interpret the linking numbers $L_{ij} \neq 0 $ as mutual inductance (usually denoted as $M$~\cite{TN_libero_mab214116274}) of coupled coils, with $L_{ij} = 0 $ corresponding to completely uncoupled elements. 
Note, that one may need to call it mutual inductance rate actually for the analogy's sake, as it would need to correspond to the change in current for further circuit comparisons.
With that in mind, the vector
\alg{
\vc{u}_1=\vc{L}\vc{M}_2^{\dagger}\vc{p}_2\\
\vc{u}_2=\vc{L}^T\vc{M}_1^{\dagger}\vc{p}_1
}
would correspond to the loop voltage vector (via mutual inductance of coupled coils). \\
So how does this help us understand $\vc{S}_i$?
Well, one may interpret any scalar ${P_i = \vc{p}_i^T \cdot \vc{S}_i \cdot \vc{p}_i}$ as a power dissipation in the circuit analogy as we may write
\alg{
	P_1 = \vc{p}_1^T \cdot \vc{S}_1 \cdot\vc{p}_1 = \vc{u}_2^T \cdot \vc{Z}_2^{-1} \cdot \vc{u}_2\\
	P_2 = \vc{p}_2^T \cdot \vc{S}_2 \cdot\vc{p}_2 = \vc{u}_1^T \cdot \vc{Z}_1^{-1} \cdot \vc{u}_1
}
Note that this would denote the power dissipation of currents in the respective partner networks.
Therefore we expect diagonal elements of $\vc{S}_i$ (corresponding to graph edges) to decay to zero if:
\begin{enumerate}
	\item the respective edges stop being a part of a coupled loop, e.g. as part of a translation of the graphs, altering topological linkage.
\item number of constituting edges in the loops (thereby the impedance) in the partner network is increased.  Be aware though that in this case we scale the diagonal values of all edges simultaneously, will not change the ranking of edge priorities.
\end{enumerate}
Be ware that changing the number of edges in an essential path(considering its linkage) of a network may change the overall ranking of edge priorities for this particular network though and therefore lead to different optimal cuts found (see custom notebook examples for more detail as part of the open-source solution SnarlPy \cite{snarlpy}).
\FloatBarrier

\section{Analyzing model systems I - Intertwined vasculature in liver acinus}
 \label{sec:appendixB}
\begin{figure}[h]
  \vspace{-0.5cm} 
\subfloat[]{\includegraphics[scale=.4]{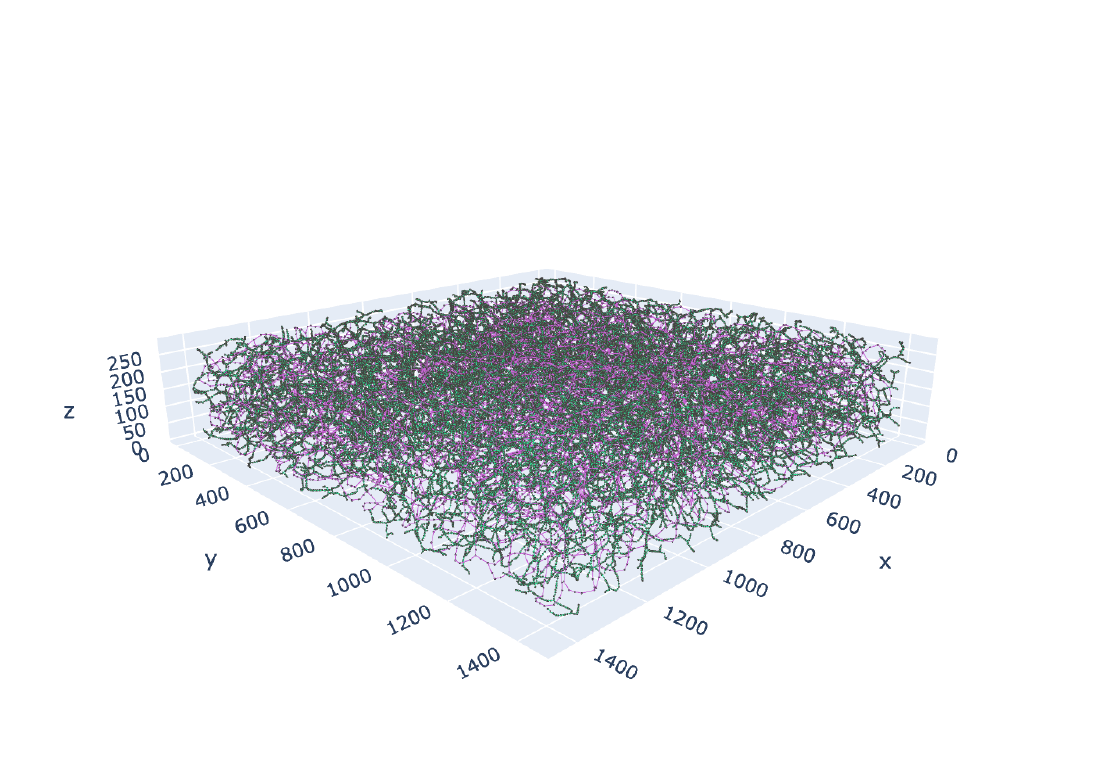}
\label{fig:classificationLiver}}
\subfloat[$\chi \in [0,150)$]{
\label{fig:classificationLiver0}
\hspace{-.8cm}
	\includegraphics[scale=.275]{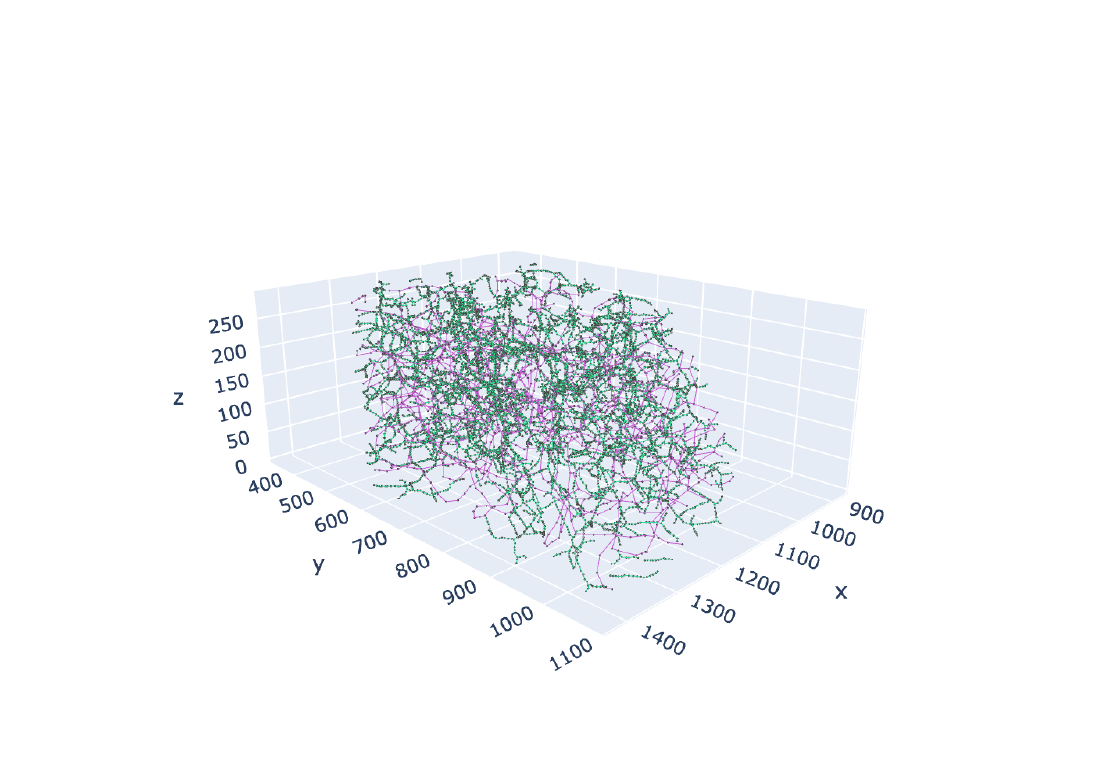}
	}
	\subfloat[$\chi \in [150,300)$]{
	\hspace{-1.1cm}
	\includegraphics[scale=.275]{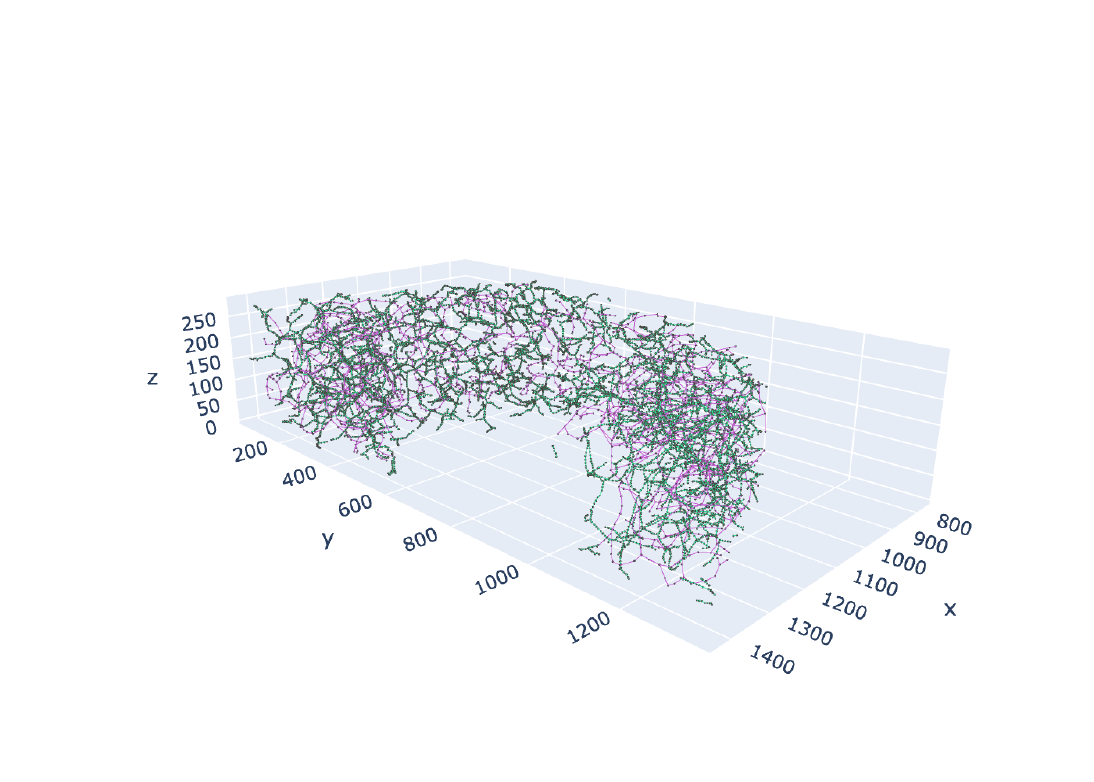}
	}\\ \vspace{-0.35cm}
	\subfloat[$\chi \in [300,450)$]{
	\hspace{-.8cm}
	\includegraphics[scale=.275]{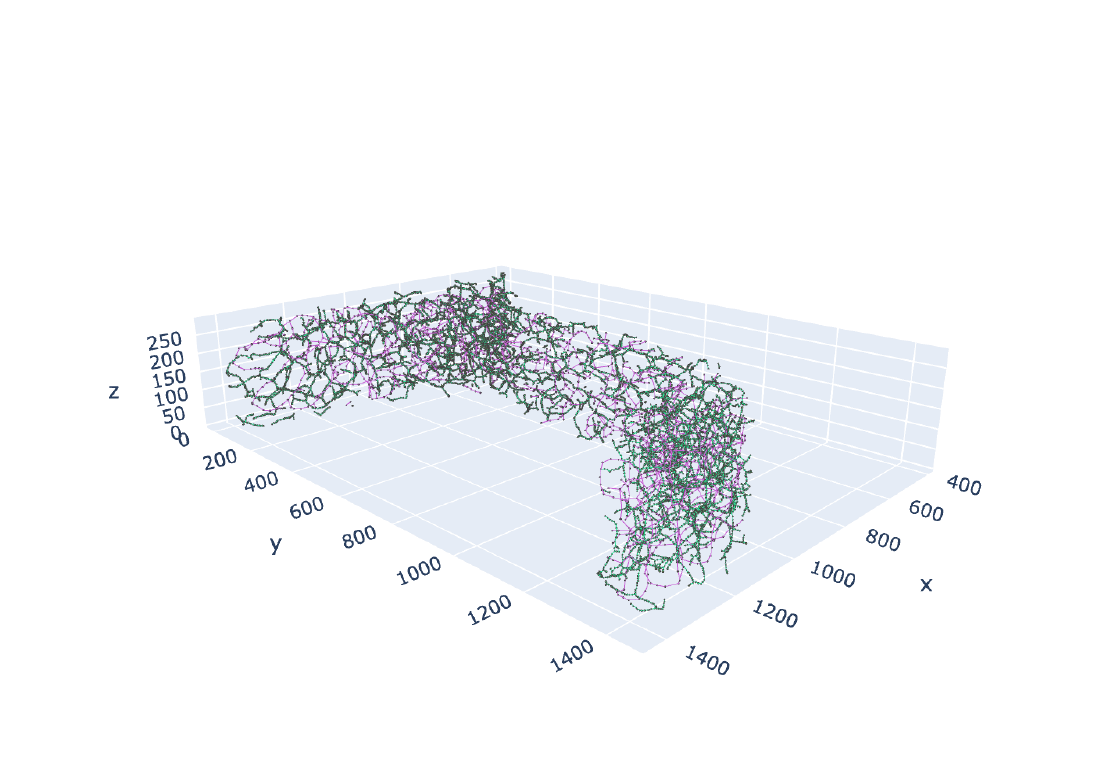}
	}
	\subfloat[$\chi \in [450,600)$]{
	\hspace{-1.1cm}
	\includegraphics[scale=.275]{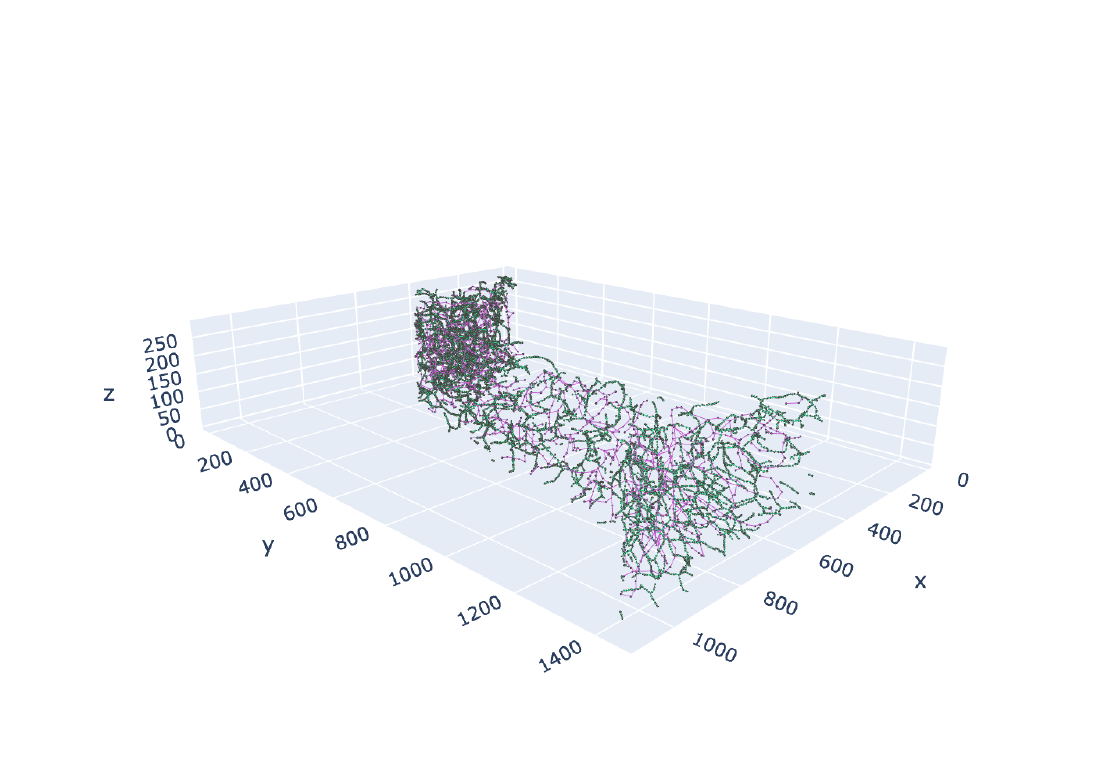}
	}
	\subfloat[$\chi \in [600,750)$]{
	\hspace{-.8cm}
	\includegraphics[scale=.275]{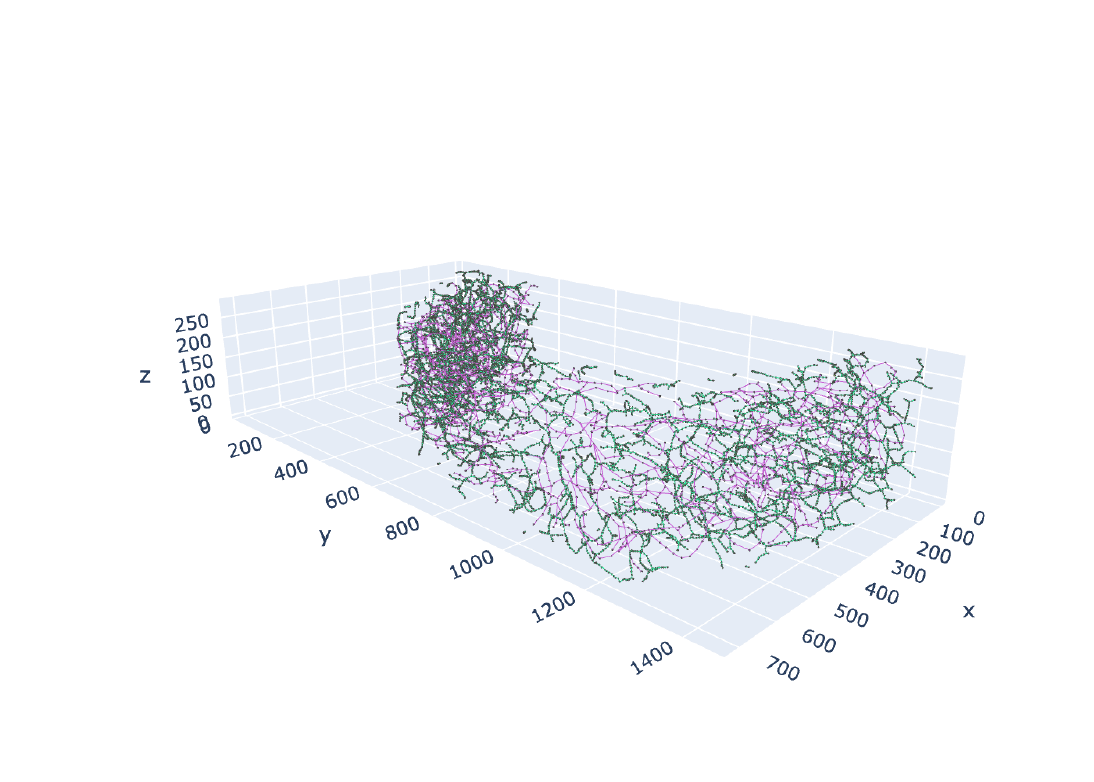}
	}
	\subfloat[$\chi \in [750,900)$]{
	\label{fig:classificationLiver1}
	\hspace{-1.1cm}
	\includegraphics[scale=.275]{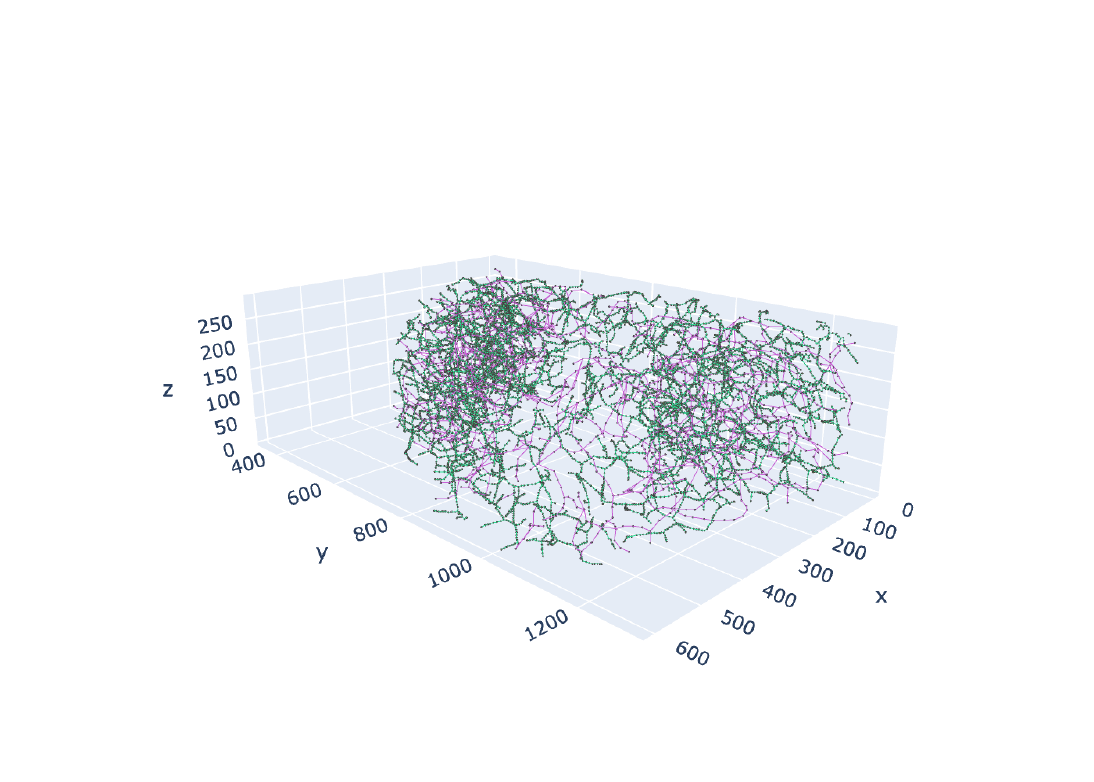}
	}
	\caption{
	Network skeletons of liver tissue vasculature:
	\subR{fig:classificationLiver} Sinusoids (magenta) and bile canaliculi (green) segment in the liver acinus.
\subR{fig:classificationLiver0}-\subR{fig:classificationLiver1} Separation of the vasculature mesh into regular $\Delta\chi=150$ sections.
	}\label{fig:classificationLiverZonation}
\end{figure}

\begin{figure}[h]
\centering
\subfloat[]{
	\includegraphics[scale=1.]{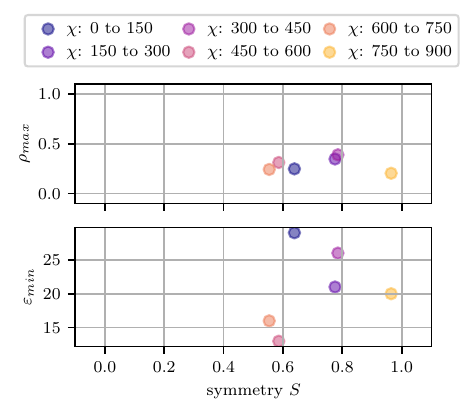}\label{fig:classificationLiverA}
	}
	\subfloat[]{
	\includegraphics[scale=1.]{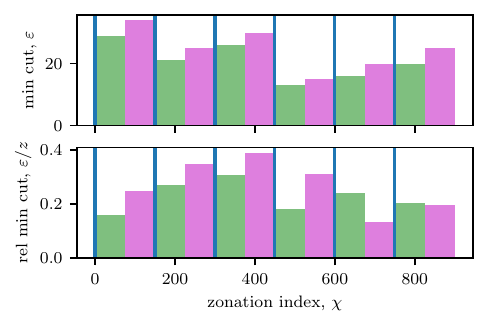}\label{fig:classificationLiverB}
	}
	\caption{
	Classification of entangled systems in the liver lobule for $\Delta\chi=150$ compartmentalization:
	\subR{fig:classificationLiverA} Minimal cut numbers and linkage parameters for liver network segments in the acinus.
	\subR{fig:classificationLiverB} Zonation and network differentiated overview of minimal cut numbers and cycle space utilization
	}\label{fig:classificationLiverOverview}
\end{figure}

 The characterization techniques presented in this study were, partially, inspired by biological model systems, mainly those found in the liver lobule (one of the fundamental tissue elements in the liver~\cite{RN152}).
 Here, sinusoids (blood vessels, endothelial tissue) and bile canaliculi (ducts for waste removal and digestive components, epithelial tissue) form an intricate, yet spatially separate, network system displaying a manifold of interesting features for a transport system~\cite{RN140,RN167,RN175}.
 Utilizing a data set from previous analysis~\cite{RN168}, we focus on a small section of tissue found in the liver acinus, an element between the central vein and a portal triad, see Figure~\ref{fig:classificationLiver}.
 Due to the segmentation techniques of the collaborators (Zerial Lab, MPI-CBG Dresden) who provided us with these data, we further have the relative positional information of vessel pieces in the acinus~\cite{RN26}, i.e. we know which part of the network is close to the central vein (denoted  as $\chi=0$) and which is most distant ($\chi=1000$).
 As it is usually of interest how the properties of liver tissue change with regard to its relative distance to the central vein/ portal triad, we perform an analysis of the linkage behavior as follows:
 We compartmentalize sections of the mesh system with regard to $\LP\chi = 150$ intervals, see Figure~\ref{fig:classificationLiver0} to~\ref{fig:classificationLiver1}.
 Subsequently we analyze the linkage of the partial mesh systems in accordance to minimal cut numbers $\varepsilon_{min}$, symmetries $S$ and effective cycle utilization $\rho_{max}$.
 Needless to say , interval size of $\LP \chi$ will influence the detected linkage behavior, as well as the generation of artifacts (say the creation of multiple connected components, and broken loops).
 We shall focus here on the largest connected component and prune it (remove all leaves) before further analysis.
 In Figure~\ref{fig:classificationLiverOverview} we display the results for zonation differentiated linkage.
 One may see that this system shows a wide range of linkage symmetry $S$ throughout the entire acinus, with severe fluctuations towards the central vein.
 Furthermore it seems one observes a rather constant overall utilization of the cycle space $\rho_{12}$ throughout the acinus, which is interesting as the minimal cut numbers decrease significantly when getting closer to the central vein.
The proposed cutting algorithm and priority computation indicates a useful tool for characterization of zonated ensnarledness, yet we find that further, more comprehensive studies are in order.
\FloatBarrier
\section{Analyzing model systems II - Optimized transport networks}
 \label{sec:appendixC}
 \begin{figure*}[t]
\centering
\subfloat[]{
	\includegraphics[scale=1.]{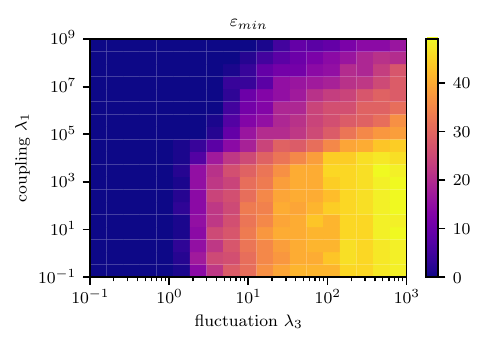}
	\includegraphics[scale=1.]{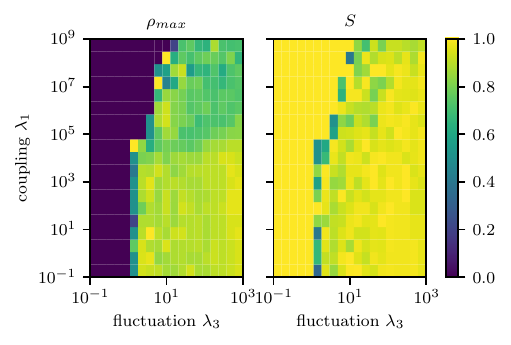}\label{fig:classificationAdaptiveRep}
	}
\\
\subfloat[]{
	\includegraphics[scale=1.]{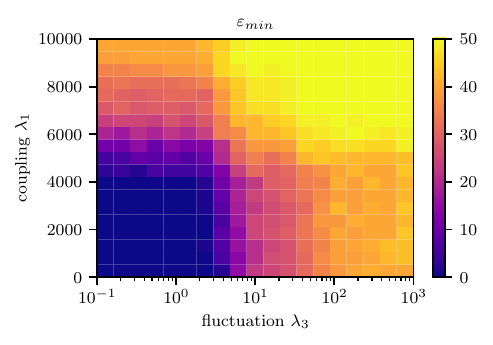}
	\includegraphics[scale=1.]{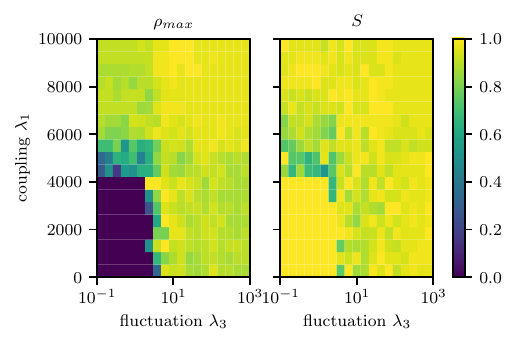}
	\label{fig:classificationAdaptiveAtt}
	}
	\caption{
	Classification of entangled, adapting and \subR{fig:classificationAdaptiveRep} repulsively coupled systems ($\nu > 1$) as we all \subR{fig:classificationAdaptiveAtt} attractively coupled systems ($\nu < -1$) systems. 
	}
\end{figure*} 
 In this section we present a linkage analysis of intertwined adapting, insilico transport networks, which represent another toy model used for capillary systems \cite{RN168}.
The transport networks of interest are modeled as Kirchhoff networks, i.e. allowing us to utilize a lumped parameter model with flows $f$ as direct linear response to potential differences $\LP p$, e.g. a Hagen-Poiseuille law \cite{RN347} $f_e=\frac{\pi r_e^4}{8\eta_e}\frac{\Delta p_e}{l_e}$ and $\sum_e B_{en} f_e = s_n$ for all edges $e$ and nodes $n$. 
Here edges in the network are assigned a length $l$, a local fluid viscosity $\eta$ and vessel radii $r$ and the graphs incidence matrix $B_{en}$.
In \cite{RN168} one defines a multilayer network consisting of two ensnarled, yet spatially separate meshes. 
Here we perform all simulation on 'laves3' systems, i.e. see the mesh 'lavesV1' in Figure 5 (main article), with sources placed in spatially opposing corners of the networks.
This model incorporates affiliation of edges: Given a minimal cycle (no chords or further deconstruction into smaller cycles possible) in a network we say that all its composing edges are affiliated with edges of other other cycles linked with the former one.
As all edges are simply tubes in our model, we have the distance between affiliated tubes defined as,
		\g{\LP r_{ee'}=L-\bra{r_e+r_{e'}}}
where L is the initial distance of the skeletons (equal to distance in case of simultaneously vanishing radii). Hence one describes this relationship between the edges $e$, $e'$ as
		\alg{
		 F_{ee'} & =\begin{cases}
1 & \text{if edges }e \text{ and }e' \text{affiliated}\\
0& \text{else}
\end{cases}}
As discussed in \cite{RN159} one finds sink-source fluctuations to be a crucial component in transport network adaption
 We only consider $\vc{s}$-configurations in which there exists one source-node (here $j=0$) and all other nodes being sinks with the following characteristics:
		\g{\mean{s_j}=\mu\text{ with }j>0\\ \mean{s_j s_k}=q_{jk}\sigma^2+\mu^2 \text{ with } j,k > 0 \label{eq:sink_condition}}
Hence one may compute the effective pressure response $\mean{\LP p_e}$ for any individual edge. 
Eventually we utilize an adaption scheme of edge radii in either network in response to the fluid flow, volume penalties and affiliation (modeled as power law with exponent $\nu$ feedback as
	  \alg{
	 \partd{h}r_e & \propto \brc{\mean{\LP p^2_e}r_e^2-\alpha}r_e-\beta\sum_{e'}F_{ee'}\LP r_{ee'}^{-\bra{\nu+1}} \label{eq:entangled_0}
	  }
 with auxiliary coefficients $\chi$, $\alpha$, $\beta> 0$. 
For numerical evaluation we use the unit and parameter system system of \cite{RN168}: 
radii $r_e=L r^*_e $, sink fluctuation $s_n=\mu \varsigma_n$, the conductivity $k_e=\eta^{-1}L^4 \kappa_e$ and hence pressure $\LP p_e=\frac{\mu\eta}{L^4} \LP\Phi_e$ and the networks' edge surface distance $\LP r_{ee'}^{-\bra{\nu+1}}=L^{-\bra{\nu+1}}\LP r_{ee'}^{-\bra{\nu+1}*}$. 
The ODEs \eqref{eq:entangled_0} can now be rescaled accordingly to the sink mean $\mu$, providing us with
 with the effective temporal response parameters ${\lambda_0=\chi\bra{\frac{\mu\eta}{L^3}}^2}$, the effective network coupling ${\lambda_1=\frac{\beta}{\pi \eta L^{\nu+1}}\bra{\frac{\mu}{L^3}}^{-2}}$, the effective volume penalty ${\lambda_2=\frac{4\alpha}{\pi\eta}\bra{\frac{\mu}{L^3}}^{-2}}$and the effective flow-fluctuation ${\lambda_3=\frac{\sigma^2}{\mu^2}}$. 
The coefficients $\lambda_2$, $\lambda_2'$ are generally negligible as shown in \cite{RN10,RN168}.
In Figure \ref{fig:classificationAdaptiveRep} and \ref{fig:classificationAdaptiveAtt} we present the result for systematic parameter scans of affiliation coupling $\lambda_1$ and fluctuation $\lambda_3$ in the case of repulsive ($\nu > 1$) and attractive interactions ($\nu < 0$).
Here we present the state diagrams for minimal cut numbers $\varepsilon_{12}$, symmetry $S$ and utilization  $\rho_{12}$. 
Generally one finds the outline of the linkage transition to match the nullity transition discussed in \cite{RN168}. 
Note that the adaptation parameters $\lambda_i$ are chosen identically for all involved systems.
It is interesting to note that the symmetry coefficient $S$ seems particular sensitive with regard to the regimes of the nullity transition, indicating different rates of acquiring reticulation in the networks probed. 
We assume this most likely to be an artifact of the chosen embedding though.
Eventually, we would like to point out that the metric is applicable to even more sophisticated systems, e.g. ensnarled networks which involve stimuli such as metabolite uptake~\cite{kramer2022}.
\FloatBarrier

\bibliography{./lib/references}